\documentclass[fleqn,usenatbib,useAMS]{mnras}

\usepackage{comment}
\usepackage[T1]{fontenc}
\usepackage{ae,aecompl}
\DeclareRobustCommand{\VAN}[3]{#2}
\let\VANthebibliography\thebibliography
\def\thebibliography{\DeclareRobustCommand{\VAN}[3]{##3}\VANthebibliography}

\usepackage{graphicx}	
\usepackage{amsmath}	
\usepackage{amssymb}	
\usepackage{subcaption}  
\usepackage{threeparttable}  
\usepackage{float}
\usepackage{multicol}        
\usepackage{newtxtext,newtxmath}

\title[BLR clouds in an AGN sample]{Broad-line Region Clouds orbiting an AGN sample}

\author[Armijos-Abenda\~no et al.]{
Armijos-Abenda\~no J.,$^{1}$\thanks{E-mail: jairo.armijos@epn.edu.ec}
L\'opez E.,$^{1,2}$
Llerena M.,$^{3}$
and Logan C. H. A.$^{4}$
\\
$^{1}$Observatorio Astron\'omico de Quito, Escuela Polit\'ecnica Nacional, 170136, Quito, Ecuador\\
$^{2}$Facultad de Ciencia, Departamento de F\'isica, Escuela Polit\'ecnica Nacional, 170136, Quito, Ecuador\\
$^{3}$Departamento de F\'isica y Astronom\'ia, Universidad de La Serena, Av. Juan Cisternas 1200 Norte, La Serena, Chile\\
$^{4}$HH Wills Physics Laboratory, University of Bristol, Tyndall Avenue, Bristol, BS8 1TL, UK
}

\date{Accepted XXX. Received YYY; in original form ZZZ}

\pubyear{2021}

\begin{document}
\label{firstpage}
\pagerange{\pageref{firstpage}--\pageref{lastpage}}
\maketitle

\begin{abstract}
We present a spectral and temporal analysis of \textit{XMM-Newton} data from a sample of six galaxies (NGC 3783, Mrk 279, Mrk 766, NGC 3227, NGC 7314, and NGC 3516). Using the hardness-ratio curves, we identify time-intervals in which clouds are eclipsing the central X-ray source in five of the six sources. We detect three occultations in NGC 3227 and one occultation in NGC 3783, NGC 7314, and NGC 3516, together with the well-known occultations in Mrk 766.
We estimate the physical properties of the eclipsing clouds. The derived physical size of the X-ray sources ($\sim$(3-28)$\times$10$^{13}$ cm) is less than that of the eclipsing clouds with column densities of $\sim$10$^{22}$-10$^{23}$ cm$^{-2}$, thus a single cloud may block the X-ray source, leading to notorious temporal variability of the X-ray flux. The eclipsing clouds in Mrk 766, NGC 3227, NGC 7314, and NGC 3516 with distances from the X-ray source of $\sim$(0.3-3.6)$\times$10$^4$ $R_g$ are moving at Keplerian velocities $>$1122 km s$^{-1}$, typical parameters of broad-line region clouds, while the eclipsing cloud in NGC 3783 is likely located in the dusty torus. We also find a good anti-correlation with a slope of -187$\pm$62 between the known masses of the supermassive black hole in the center of the galaxies with the equivalent width (EW) of the 6.4 keV Fe line for the five Seyfert 1 galaxies of our sample, while the Seyfert 2 galaxy NGC 7314 shows an average EW value of 100$\pm$11 eV inconsistent with the above anti-correlation.

\end{abstract}




\section{Introduction}

Absorption-driven X-ray variability detected in active galactic nuclei (AGNs) has been interpreted as the passage of absorbing clouds, occulting the central X-ray source (e.g., \cite{Risaliti2011,2014MNRAS.442.2116T}). Absorption features are expected in the ultraviolet and X-ray spectra when a large cloud or a group of small clouds crosses our line of sight \citep{2009ApJ...695..781B,2012MNRAS.424.2255W}. The absorbing clouds seem to have properties (locations and scales) similar to those of the emitting clouds of the broad-line emission region (BLR), so the ultraviolet and X-ray absorbing clouds may be the clouds known as the BLR clouds.
Strong absorption-driven variability is expected when low ionization absorbers with neutral cores are eclipsing the central X-ray source \citep{Risaliti2011,2009NewAR..53..140G}. In this case, the iron absorption lines are physically associated with the absorber \citep{Risaliti2011}.
The BLR, with a size between $\sim$10$^{-4}$ to 0.1 pc \citep{2019A&A...628A..26P}, is illuminated by the photo-ionizing continuum radiation of the AGN, reprocessing it into
emission lines \citep{2009NewAR..53..140G}.
The cloud geometry is unknown, but it is expected to be irregular. The cloud sizes, determined by different methods, are of the order of $10^{12}-10^{14}$ cm \citep{Risaliti2011,2019A&A...628A..26P}. It has been proposed that the BLR gas may be moving on inflowing or outflowing elliptical trajectories \citep{2018ApJ...866...75W}. \cite{2015MNRAS.446.1848K} showed that the BLR clouds may be distributed in a disc-like configuration in AGNs. There is also a hypothesis that the BLR is formed by dusty clouds transferred to regions above the plane of the accretion disk by radiation pressure \citep{2011A&A...525L...8C}.

The column density $N_H$ and the covering factor $C_F$ of the clouds vary proportionally during the occultation, so that an increment of these  parameters indicates the occurrence of an occultation of the X-ray source by a cloud (or clouds), which is absorbing/reflecting radiation \citep{2014MNRAS.442.2116T}. Therefore, the absorption-driven X-ray variability of the source is interpreted as the change of the absorbing column density along the line of sight due to occultation by clouds in the BLR, with about the same size as the X-ray source \citep{Risaliti2009}. It is acceptable to think that the coverage time of the source depends on the cloud size and its velocity. Occultation events in AGNs together with the geometrical and kinematical parameters of eclipsing clouds have been studied before (e.g., \cite{Risaliti2009,Risaliti2011,2014MNRAS.442.2116T}).

BLR models describe the geometry and physical conditions in the line emitting region of AGNs \citep{2017ApJ...847...56E,2018ApJ...866...75W,2020MNRAS.492.5540M}. The broad emission lines may originate in cool clouds ($T\sim 10^4-10^5$ K, e.g. \cite{1990ApJ...352..423M,1996MNRAS.283.1322K}). The cool clouds are considered magnetically confined and in thermal equilibrium, in order to explain their long survival times \citep{1997MNRAS.284..717K}.

Considering the standard AGN model, it is thought that the supermassive black hole (SMBH) is surrounded by an accretion disk and a dusty torus, and that the BLR clouds move in Keplerian orbits inside the dusty torus \citep{Goad2012,Muller2020}, while NLR (narrow-line region) clouds are external to the dusty torus \citep{2015ARA&A..53..365N}. In addition, it is expected we only observe NLRs towards Seyfert 2 galaxies as the torus obscures the BLRs. On the other hand, both NLRs and BLRs would be observed toward Seyfert 1 galaxies as the system is more face-on relative to the observer.

X-ray variability of the Seyfert 1 galaxies that we will consider in our study has been studied by \cite{1999ApJ...524..667T}, who proposed that differences in the mass of the SMBH and the accretion rate could explain the X-ray variability in the studied AGNs, but differences in the physical conditions and geometry of the circumnuclear gas can also be invoked to explain the observed X-ray variability.\\

The present study aims to identify the presence of cloud occultations of the central X-ray source in six AGNs (see Table \ref{table1}) by using the hardness-ratio (HR) light curves, as well as to characterize the physical properties of the clouds eclipsing the central X-ray source. For our study, we selected five Seyfert 1 galaxies since it is more likely to observe BLRs towards this type of galaxies.
We have also included one Seyfert 2 galaxy (NGC 7314) in our sample to find possible differences in the derived parameters of the Seyfert 1 galaxies with those of the Seyfert 2 galaxy.
Here, we study the physical properties of BLR clouds following the methods used by \cite{2007ApJ...659L.111R,Risaliti2011}. As far as we are aware, these methods have not been used before to study BLR clouds of our sample of galaxies, except for Mrk 766 (included in our study to check part of our results and for comparison purposes).
\cite{2014MNRAS.439.1403M} studied obscuring clouds in NGC 3783 and NGC 3227, proposing that they are located in the dusty torus, but the method used by \cite{2014MNRAS.439.1403M} to constrain obscuring cloud locations is different from those used by \cite{2007ApJ...659L.111R,Risaliti2011}.
The objects presented in this contribution are only a small sample, which will be larger in future research. Nevertheless, we suggest the results from this work provide important new insight into the AGN obscuring phenomenon. In Section \ref{Observations}, we describe the observations and data reduction. In Section  \ref{Analysis}, we present the results of temporal and spectral analysis of the X-ray data. In Section \ref{Discussion}, we discuss the relation found between the equivalent width of the Fe K$\alpha$ line at 6.4 keV and the mass of the supermassive black hole, as well as the physical properties of the eclipsing clouds. Finally, in Section \ref{Conclusion} we report the conclusions of the present study.

\section{Observations and data reduction}\label{Observations}

\begin{figure*}
    \centering
    \includegraphics[width=0.9\textwidth]{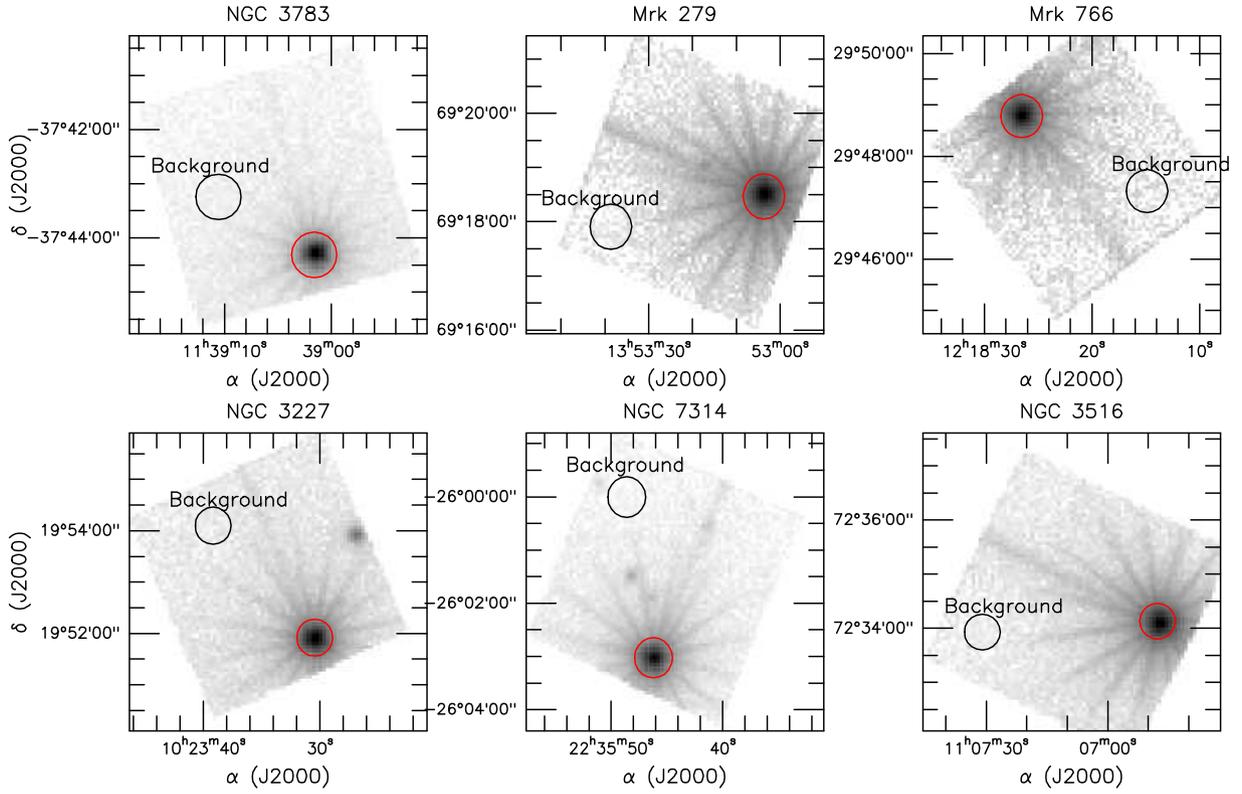}
    \caption{1-10 keV band maps of the six AGNs studied in this paper. The black circle shows the region used as background, while the red circle shows the region used to extract spectra and light curves for each galaxy.}
    \label{Sample_galaxies}
\end{figure*}

The XMM-Newton space mission has made many observations of Seyfert galaxies, and a large database is available through the XMM-Newton Science Archive\footnote{http://nxsa.esac.esa.int/nxsa-web/\#search}. We selected a sample of six galaxies from this archive with long observations times of at least $\sim$40 ks, which is necessary to see changes in the HR light curves of the galaxies and thus identify possible occultations of the central X-ray source (see below).
These galaxies are also selected for our study as the mass of the supermassive black holes is known and changes within $\sim$(1-80)$\times$10$^6$ M$_{\odot}$ \citep{Christopher2002,Onken2003,Giacche2014,Piotrovich2015,Emma2016}, which will allow us to find possible relations between the mass of supermassive black holes with derived parameters. This is a pilot project which will be extended to other galaxies with high redshift $z$, in forthcoming research.
The selected data were observed with the EPIC instrument. The Seyfert type, distance, observation ID, and the exposure time are given in Table \ref{table1}, for each galaxy. The \texttt{epproc} task of the Science Analysis Software\footnote{See \url{https://www.cosmos.esa.int/web/xmm-newton/sas}} (SAS, version 19.1.0) was used to run the default pipeline processing, thus obtaining the calibrated event lists. We then used the \texttt{evselect} task of SAS for filtering the data, as well as for extracting light curves and spectra toward the central regions of the galaxies. Fig. \ref{Sample_galaxies} shows the X-ray emission maps of the six galaxies. We selected a region free from contamination sources for the background of each galaxy. We used the same radius size of the source for the background regions.\\

We have studied the X-ray emission arising in the central circular regions with a radius of 21 arcsec for NGC 7314, 25 arcsec for NGC 3783, NGC 279, Mrk 766, and 29 arcsec for NGC 3516 (see Fig. \ref{Sample_galaxies}). These galaxies are nearby sources with redshifts within 0.004-0.031 and bright with 2-10 keV fluxes within $\sim$(1-5)$\times$10$^{-11}$ erg cm$^{-2}$ s$^{-1}$ (see below), so the above source radii can be adopted for our study. As mentioned above, five of the studied galaxies are Seyfert 1 type, so the detection of eclipsing events is expected, while the Seyfert 2 galaxy is included in this study for comparative purposes.

\begin{table}
	\centering
	\caption{Observational parameters of the studied galaxies.}
	\label{table1}
	\begin{tabular}{lcrcr} 
		\hline
		Galaxy & type & distance & obsIDs & exposure time\\
		       & & (Mpc) &  & ($\times$10$^3$ sec)\\
		\hline
		NGC 3783 & Seyfert 1 & 41.6 & 0780860901 & 115.0\\
		      & &           & 0780861001          & 57.0  \\\hline
		Mrk 279 & Seyfert 1 & 131.7 & 0302480401 & 59.8\\
		       & &     & 0302480501 & 59.8 \\
		       & &     & 0302480601 & 38.2 \\ \hline     
		Mrk 766 & Seyfert 1 & 64.9 & 0304030101 & 95.5\\
		       & &     & 0304030301 & 98.9 \\
		       & &     & 0304030401 & 98.9 \\ \hline
		NGC 3227 & Seyfert 1 & 22.4 & 0782520201 & 92.0\\
		      & & & 0782520301 & 74.0\\
		      & & & 0782520401 & 84.0\\
		      & & & 0782520501 & 87.0\\ \hline
	    NGC 7314 & Seyfert 2 & 15.8 & 0725200101 & 140.5\\
	          & & & 0725200301 & 132.1\\ \hline
	    NGC 3516 & Seyfert 1 & 38.9 & 0401210401 & 52.2\\ &  &  & 0401210501 & 69.1\\
	    &  &  & 0401210601 & 68.5\\
	    &  &  & 0401211001 & 68.6\\
		\hline
	\end{tabular}
\end{table}

\section{Analysis of the X-ray data}\label{Analysis}
\subsection{Temporal analysis}\label{Analysis1}

\begin{figure}
\centering
\includegraphics[trim=1.5cm 0 1.8cm 0,width=0.42\textwidth]{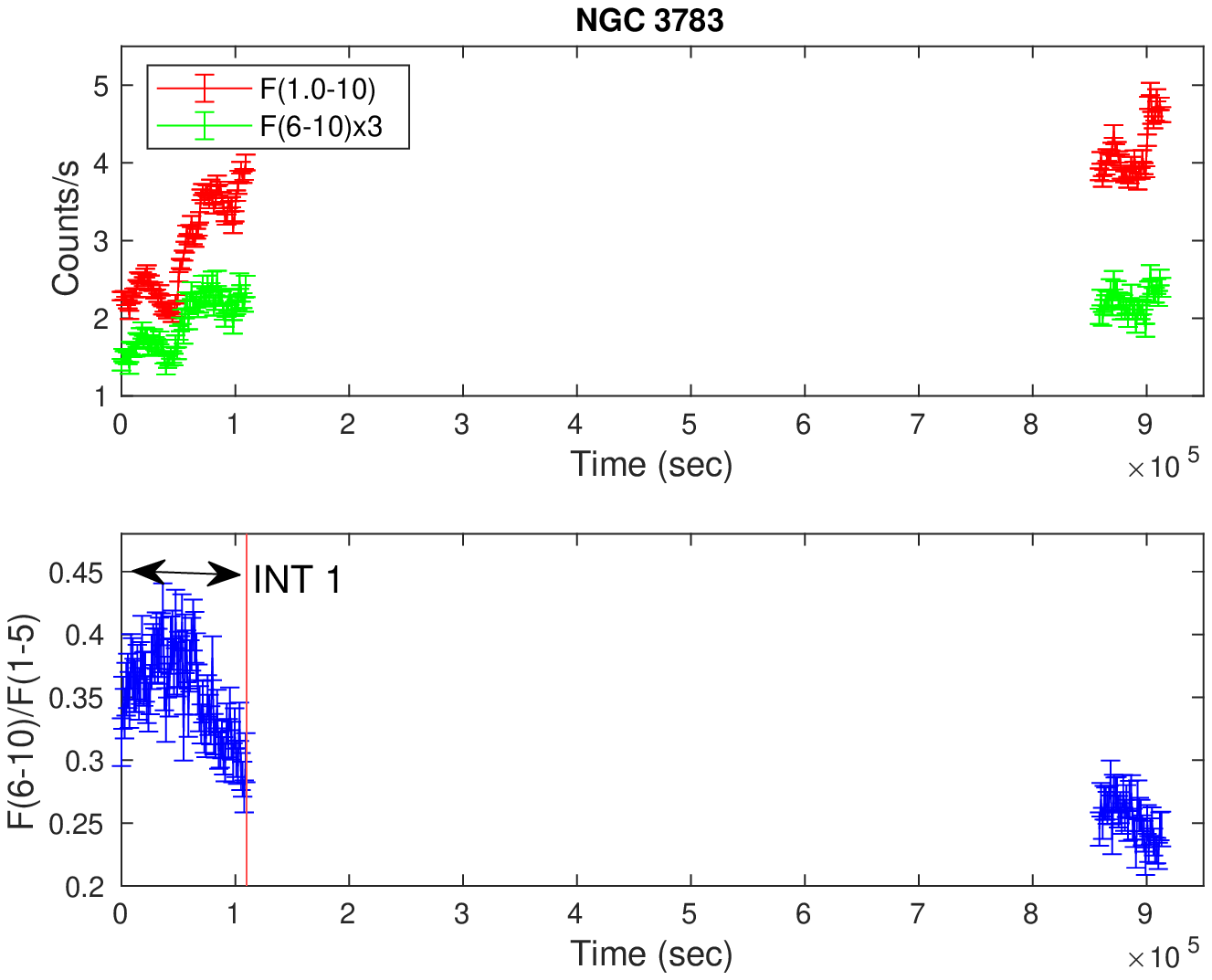}   
\caption{The 1-10 and 6-10 keV flux light curves (top panel) and hardness-ratio (bottom panel) of NGC 3783. The time-interval 1 showing a change in the hardness-ratio is indicated.}\label{HRngc3783}
\end{figure}

\begin{figure}
\centering
\includegraphics[trim=1.5cm 0 1.8cm 0,width=0.42\textwidth]{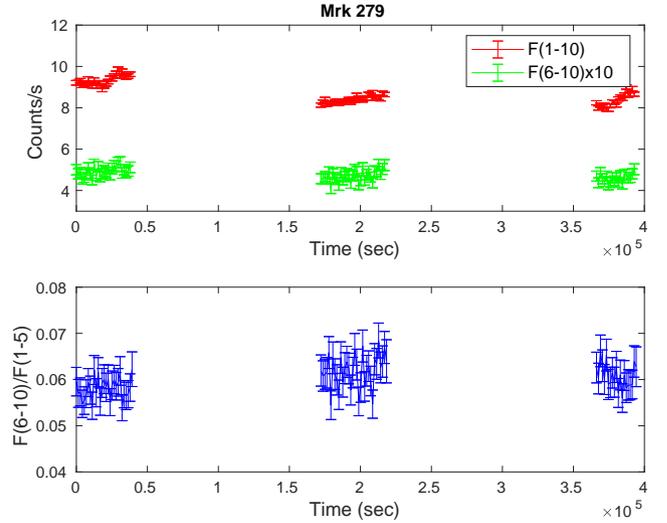}
\caption{The 1-10 and 6-10 keV flux light curves (top panel) and hardness-ratio (bottom panel) of Mrk 279.}\label{HRmrk279}
\end{figure}

\begin{figure}
\centering
\includegraphics[trim=2.7cm 0 3.5cm 0,width=0.33\textwidth]{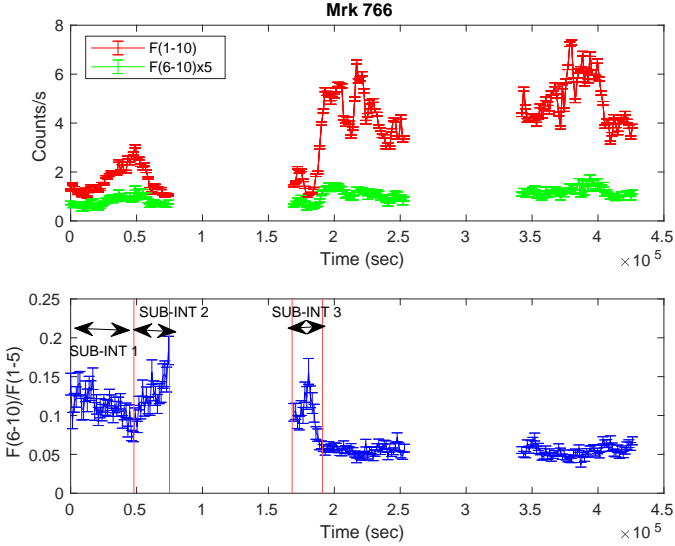}
\caption{The 1-10 and 6-10 keV flux light curves (top panel) and hardness-ratio (bottom panel) of Mrk 766. The sub-intervals 1-3 showing a change in the hardness-ratio are indicated.}\label{HRmrk766}
\end{figure}

\begin{figure}
\centering
\includegraphics[trim=4.6cm 0 4.7cm 0,width=0.44\textwidth]{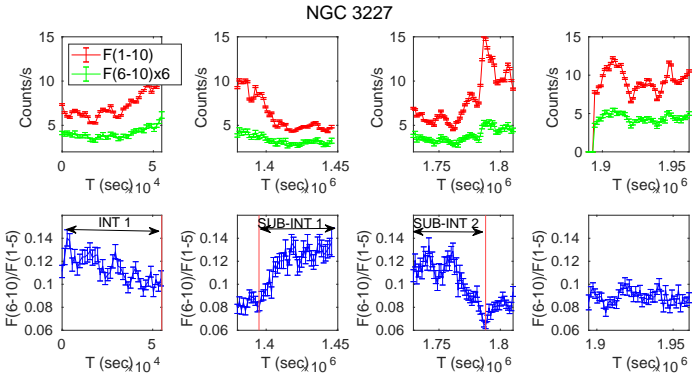}
\caption{The 1-10 and 6-10 keV flux light curves (top panels) and hardness-ratio (bottom panels) of NGC 3227. The interval 1 and sub-intervals 1-2 showing a change in the hardness-ratio are indicated.}\label{HRngc3227}
\end{figure}

\begin{figure}
\centering
\includegraphics[trim=1cm 0 1.3cm 0,width=0.47\textwidth]{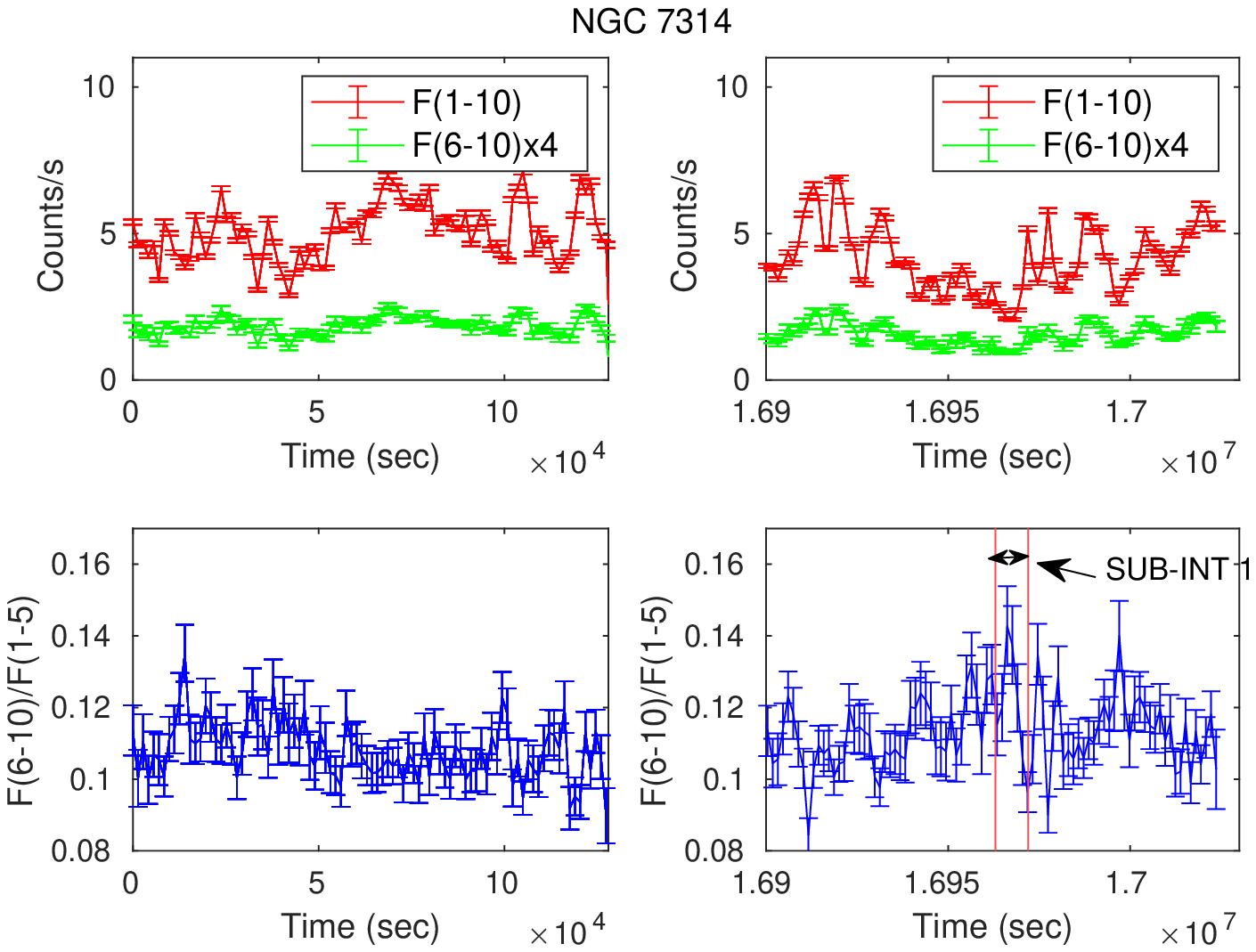}
\caption{The 1-10 and 6-10 keV flux light curves (top panels) and hardness-ratio (bottom panels) of NGC 7314. The sub-interval 1 showing a change in the hardness-ratio is indicated.}\label{HRngc7314}
\end{figure}

\begin{figure}
\centering
\includegraphics[trim=0.5cm 0 1cm 0,width=0.47\textwidth]{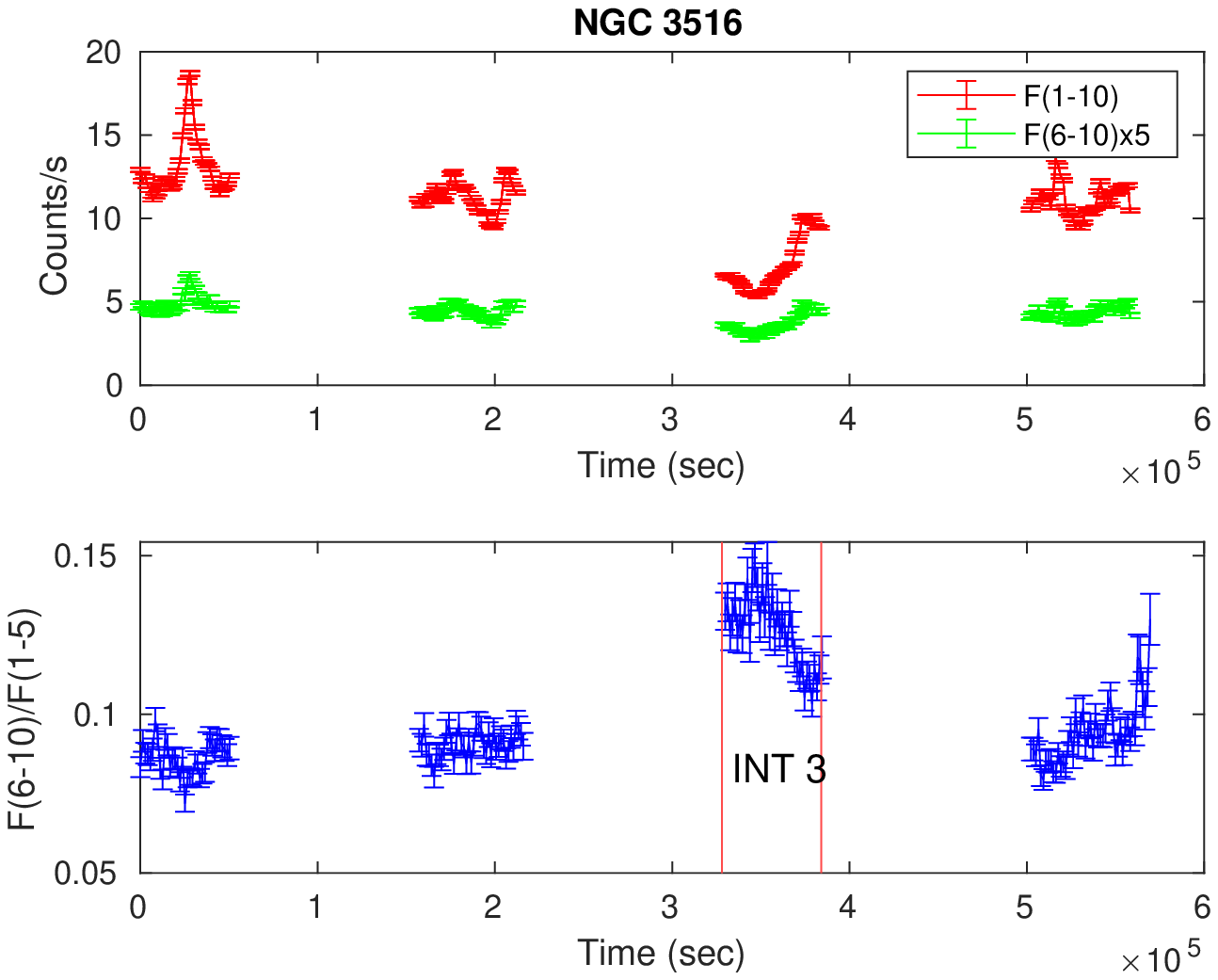}
\caption{The 1-10 and 6-10 keV flux light curves (top panels) and hardness-ratio (bottom panels) of NGC 3516. The time-interval 3 showing a change in the hardness-ratio is indicated.}\label{HRngc3516}
\end{figure}

We show the 1-10 and 6-10 keV flux light curves in Figs. \ref{HRngc3783}-\ref{HRngc3516} for the observations of the six galaxies included in our study. In these figures, we also show the hardness-ratio (HR) light curves ($F(6-10)/F(1-5)$), which reveal time-intervals with strong changes in the X-ray radiation, likely originating as a consequence of occultations of the central X-ray source by clouds crossing the line of sight. As mentioned above, for this analysis we extracted data towards the circular region of the six galaxies.\\

In Fig. \ref{HRngc3783}, the time-interval 1 shows a strong change\footnote{In this paper, we consider a change in the HR light curve when this change is greater than 25\% inside the same time-interval or there is a change in the HR light curves greater than 25\% between time-intervals.} in the HR light curve in NGC 3783. None of the studied time-intervals show changes in the HR curves in Mrk 279 (see Fig. \ref{HRmrk279}). The time-intervals 1-2 reveal strong changes in the HR curves in Mrk 766 (see Fig. \ref{HRmrk766}). We have labeled three sub time-intervals as SUB-INT 1, SUB-INT 2, and SUB-INT 3 in Fig. \ref{HRmrk766}, where changes in the HR light curves of Mrk 766 are observed. These time-intervals were studied in detail by \citet{Risaliti2011}. As mentioned above, we included Mrk 766 in our study for verifying part of the results with those of \citet{Risaliti2011}.
The time-interval 1 and sub time-intervals SUB-INT 1 and SUB-INT 2 show variations in the HR light curves in NGC 3227 (see Fig. \ref{HRngc3227}). For NGC 7314 (see Fig. \ref{HRngc7314}), the sub time-interval labeled as SUB-INT 1 shows a change in the HR curve. We notice in Fig. \ref{HRngc3516} a change in the HR light curve of the time-interval 3 in NGC 3516. In summary, NGC 3783, Mrk 766, NGC 3227, NGC 7314, and NGC 3516 show changes in their HR curves, which could be a consequence of occultations of the central X-ray source by  orbiting clouds. The HR changes evidence the possible occurrence of three cloud occulations in Mrk 766 (already discovered by \cite{Risaliti2011}) and NGC 3227, and one cloud occultation in NGC 3783, NGC 7314 and NGC 3516.\\

\begin{figure}
\begin{subfigure}{0.5\textwidth}
\centering
\includegraphics[width=0.9\textwidth]{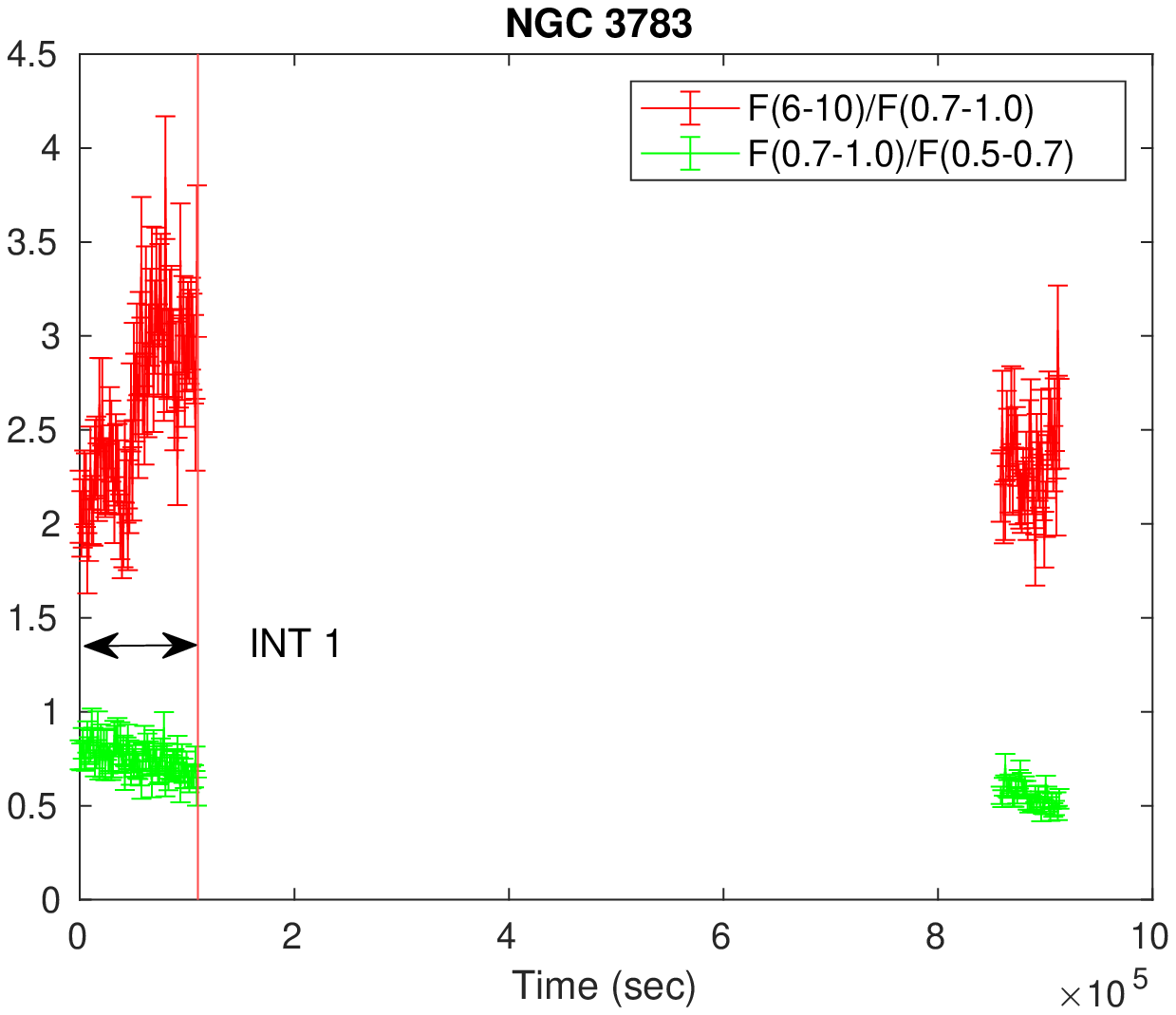}
\end{subfigure}
\begin{subfigure}{0.5\textwidth}
\centering
\includegraphics[width=0.9\textwidth]{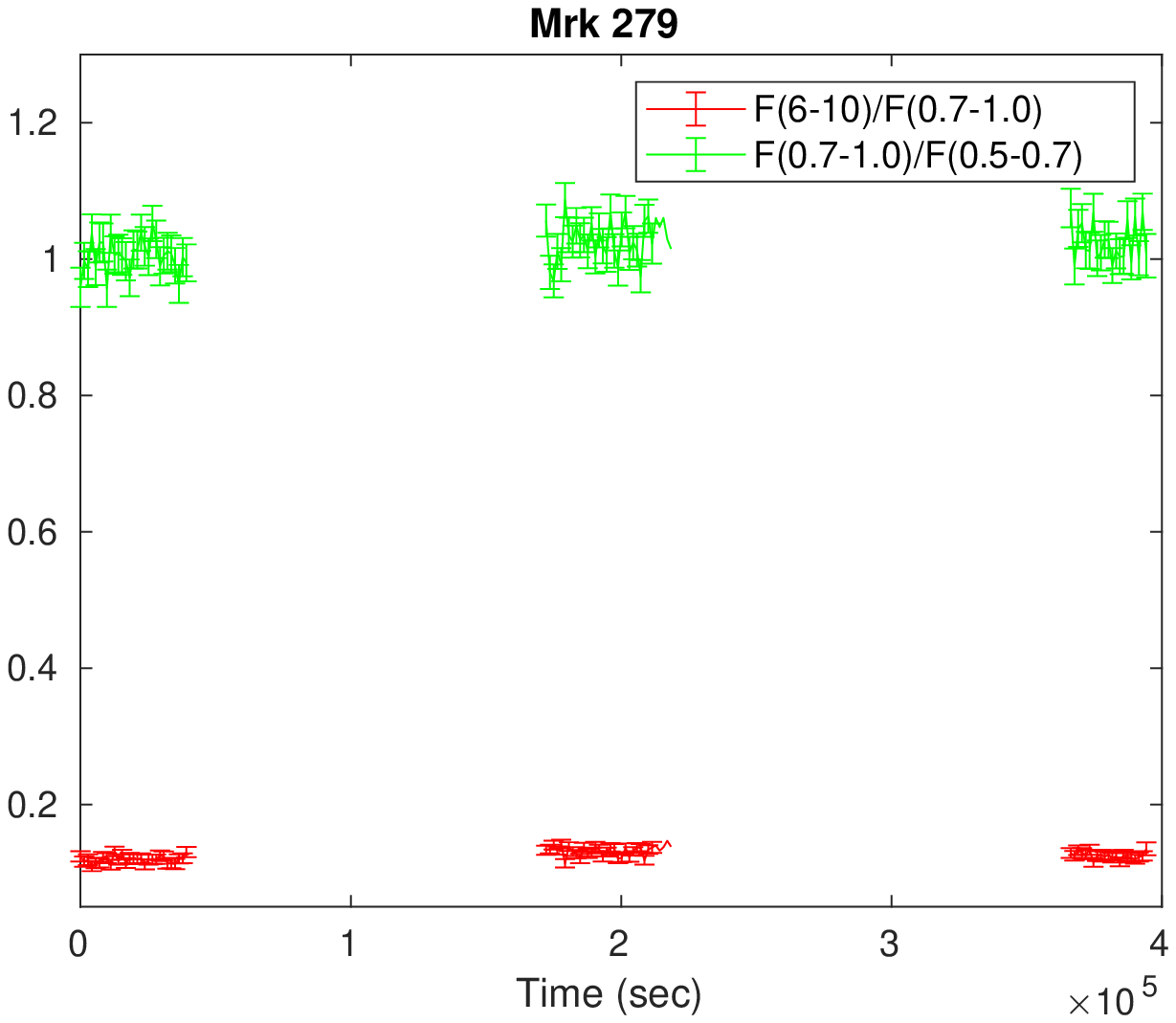}
\end{subfigure}
\begin{subfigure}{0.5\textwidth}
\centering
\includegraphics[width=0.9\textwidth]{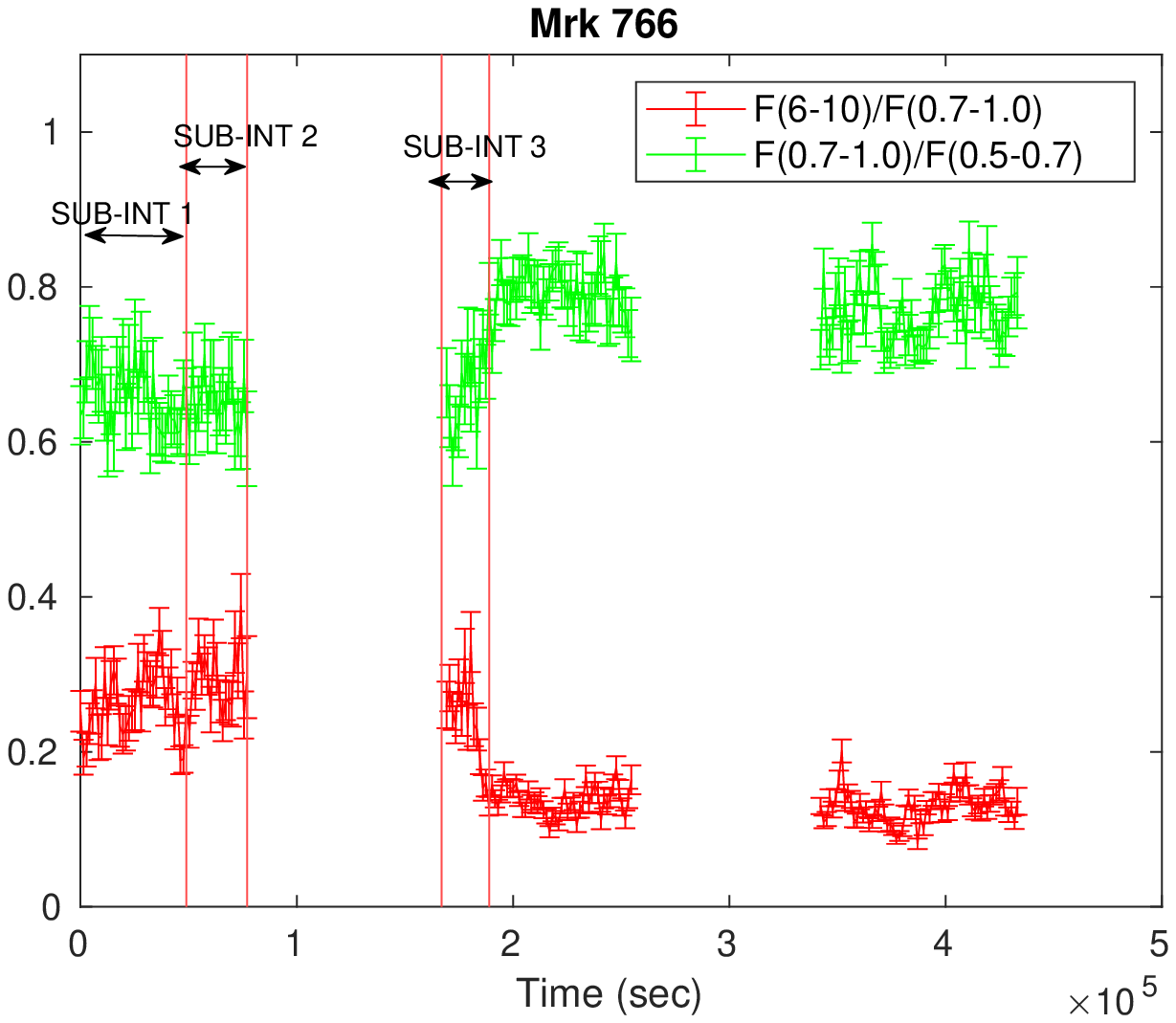}
\end{subfigure}
\caption{The curves of the 6-10 keV flux over 
the 0.7-1.0 keV flux (red) and the 0.7-1.0 keV flux over the 0.5-0.7 keV flux (green) for NGC 3783, Mrk 279, and Mrk 766 from the top to the bottom. Different intervals or sub-intervals showing a change in the hardness ratio F(6-10 keV)/F(1-5 keV) are indicated for NGC 3783 and Mrk 766.}\label{HR_galaxies1}
\end{figure}

\begin{figure}
\begin{subfigure}{0.5\textwidth}
\centering
\includegraphics[trim=3cm 0 2cm 0,width=0.9\textwidth]{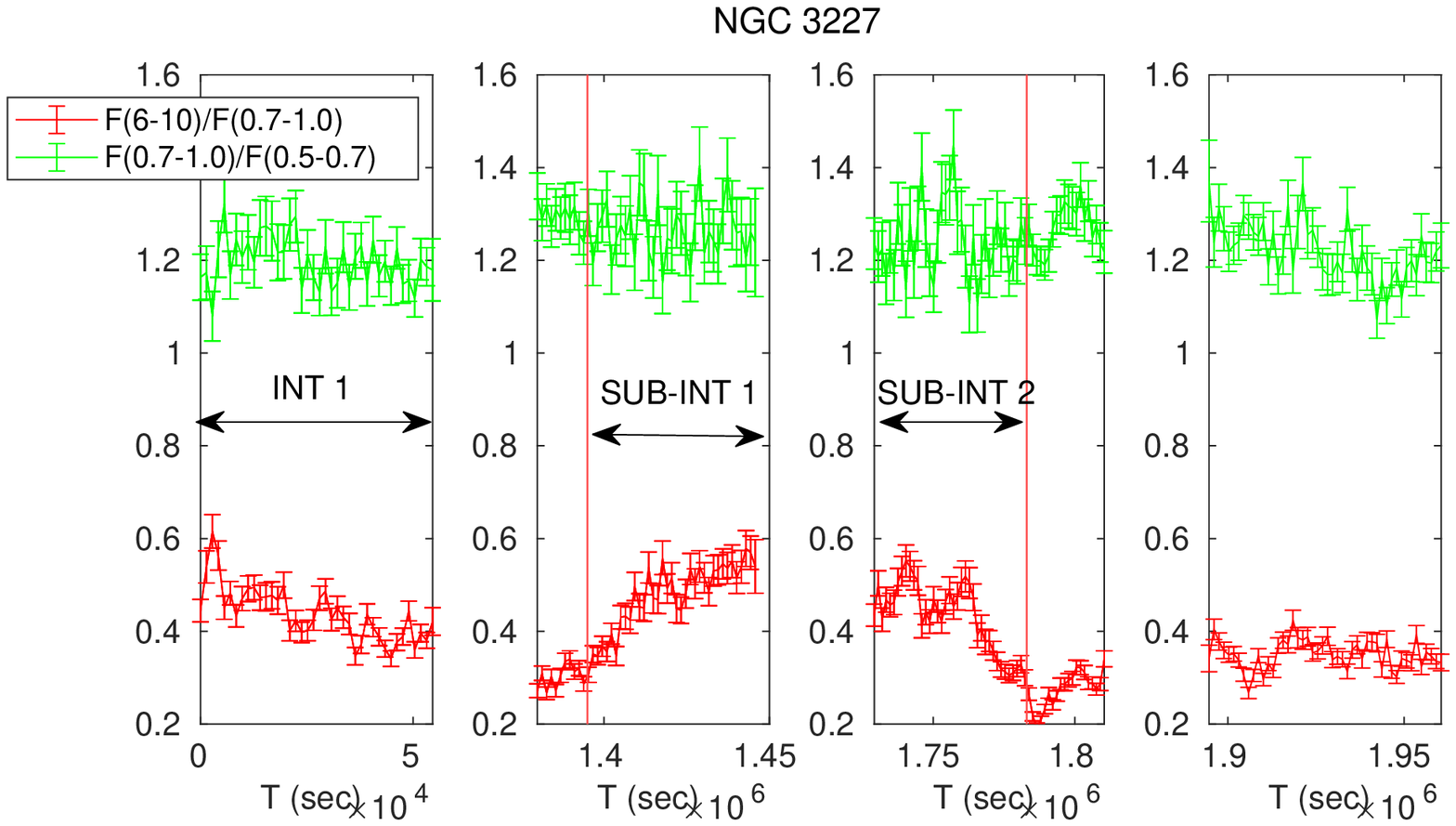}
\end{subfigure}
\begin{subfigure}{0.5\textwidth}
\centering
\includegraphics[trim=2.1cm 0 1.1cm 0,width=0.9\textwidth]{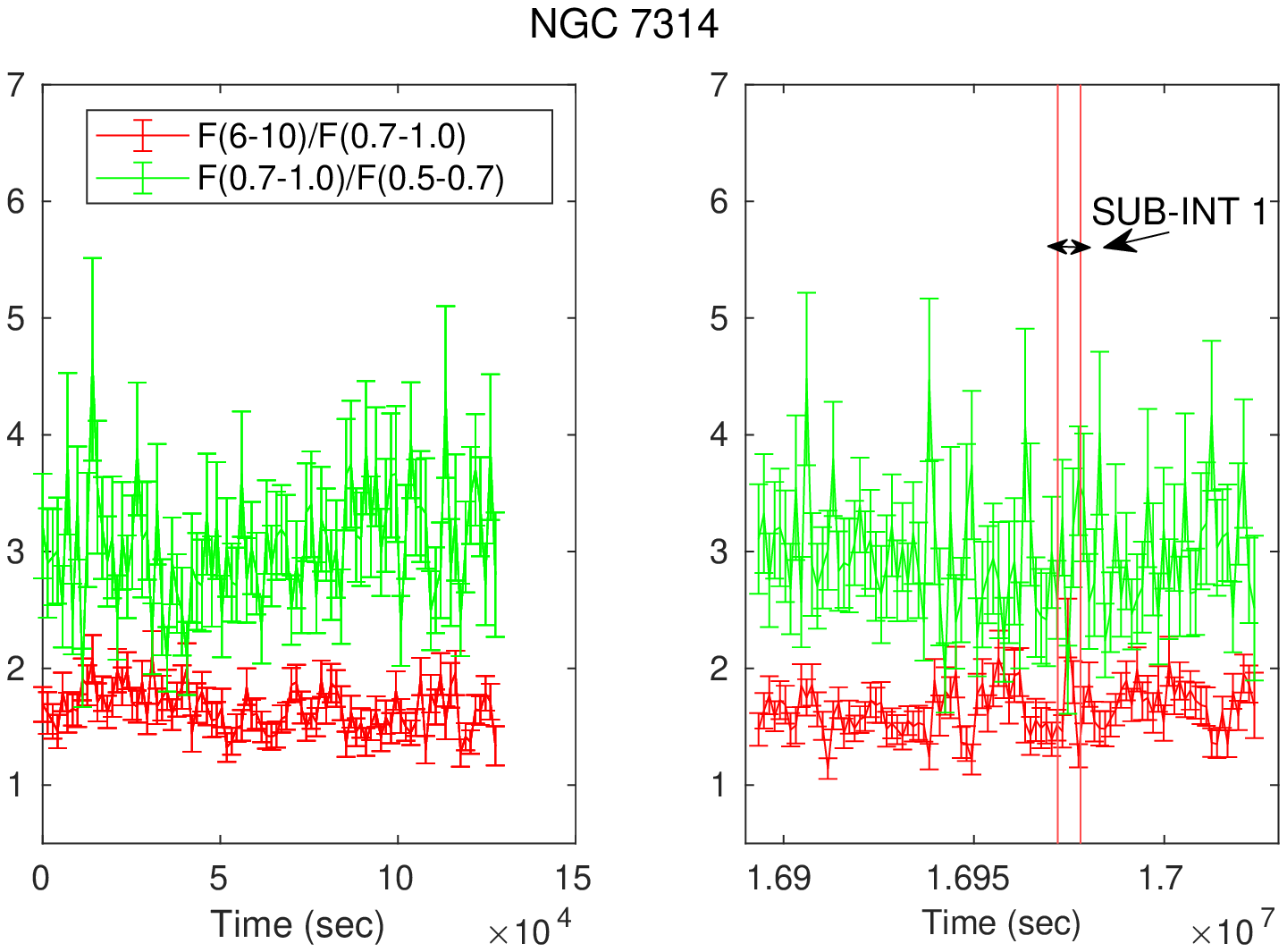}
\end{subfigure}
\begin{subfigure}{0.5\textwidth}
\centering
\includegraphics[trim=1.5cm 0 1cm 0,width=0.9\textwidth]{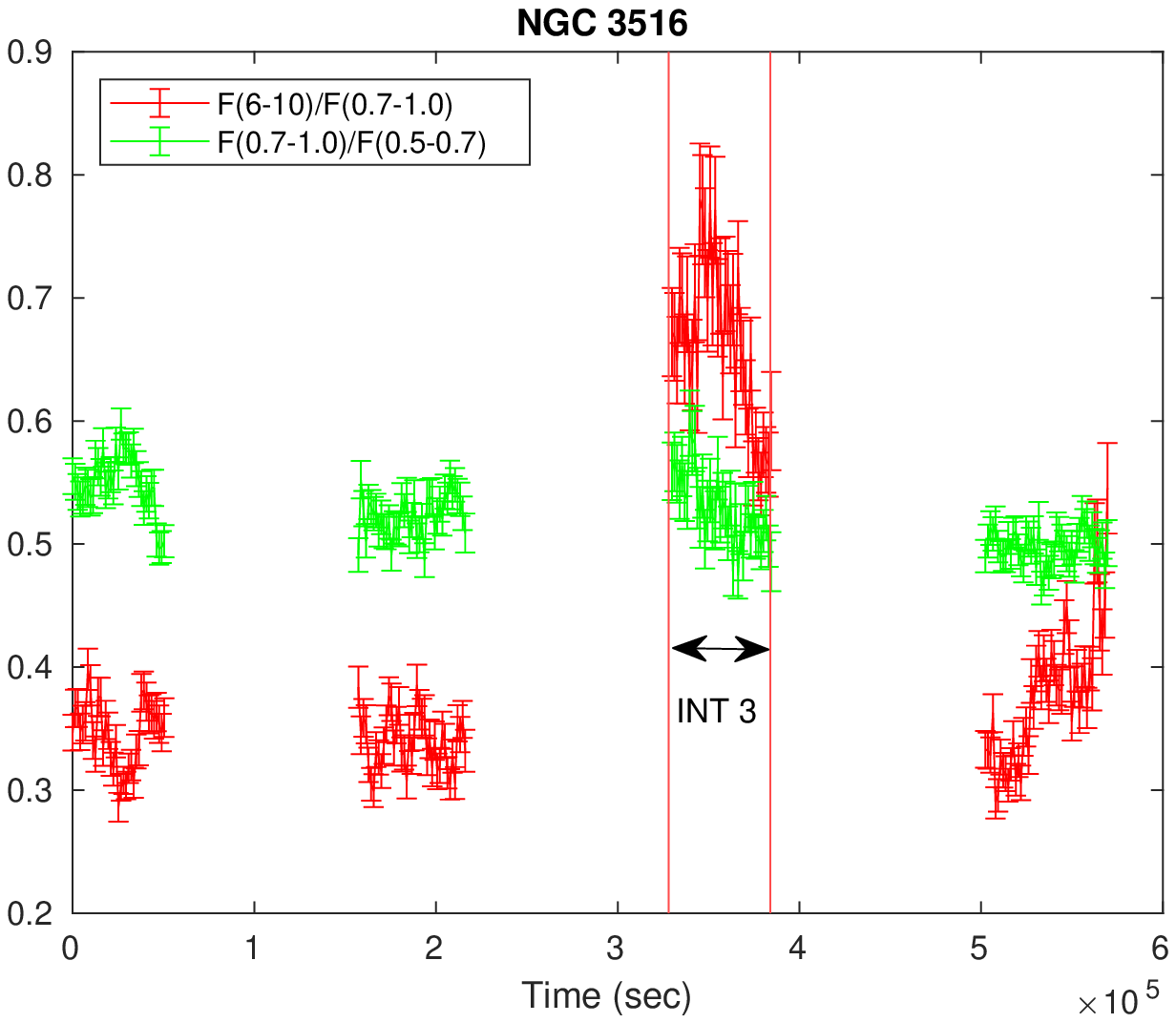}
\end{subfigure}
\caption{The curves of the 6-10 keV flux over the 0.7-1.0 keV flux (red) and the 0.7-1.0 keV flux over the 0.5-0.7 keV flux (green) for NGC 3227, NGC 7314, and NGC 3516 from the top to the bottom. Different intervals or sub-intervals showing a change in the hardness ratio F(6-10 keV)/F(1-5 keV) are indicated for NGC 3227, NGC 7314, and NGC 3516.}\label{HR_galaxies2}
\end{figure}

To see the possible effects of the soft X-ray emission (0.5-1 keV) on the ratio of the light curves, we show in Figures~\ref{HR_galaxies1} and \ref{HR_galaxies2} the F(6-10 keV)/F(0.7-1 keV) curves and the F(0.7-1 keV)/F(0.5-0.7 keV) curves. In these figures, we notice that the shape of the F(6-10 keV)/F(0.7-1 keV) curves is quite similar to that of the (6-10 keV)/F(5-1 keV) curves shown in Figures \ref{HRngc3783}-\ref{HRngc3516}. However, we find differences between the F(0.7-1 keV)/F(0.5-0.7 keV) curves and the F(6-10 keV)/F(0.7-1 keV) curves in NGC 3783, Mrk 766, NGC 3227, and NGC 3516 (see Figures \ref{HR_galaxies1} and \ref{HR_galaxies2}). 
The F(0.7-1 keV)/F(0.5-0.7 keV) curves do not show changes as those shown by the F(6-10 keV)/F(0.7-1 keV) curves in NGC 3227 and NGC 3516 (top and bottom panels in Figure \ref{HR_galaxies2}). There is an inversion in the shape of the curves of the SUB-INT 1, SUB-INT 2, and SUB-INT 3 relative to the curve of the third time interval in Mrk 766 (see the bottom panel in Figure \ref{HR_galaxies1}). The shape of the F(0.7-1 keV)/F(0.5-0.7 keV) curve in NGC 3783 is flatter than that of F(6-10 keV)/F(0.7-1 keV). There are no differences in the shape of the curves of Mrk 279 shown in the middle panel of Figure \ref{HR_galaxies1}.
The F(0.7-1 keV)/F(0.5-0.7 keV) curve is noisier than that of F(6-10 keV)/F(0.7-1 keV) in NGC 7314, which does not allow us to see the occurrence of possible eclipsing events (see the middle panel of Fig. \ref{HR_galaxies2}). The fact that the F(0.7-1 keV)/F(0.5-0.7 keV) curves do not show the occurrence of eclipsing events in NGC 3227 and NGC 3516, and that this curve is flatter than that of F(6-10 keV)/F(0.7-1 keV) in NGC 3783 suggest that the eclipsing events are best observed in the hard X-ray band (> 1 keV).

In the following section, we show a spectral analysis for the whole time-intervals, as well as a time-resolved analysis of the intervals or sub-intervals showing variations in the HR F(6-10 keV)/F(1-5 keV) curves.

\subsection{Spectral analysis}\label{Spectral_analysis}

\begin{figure*}
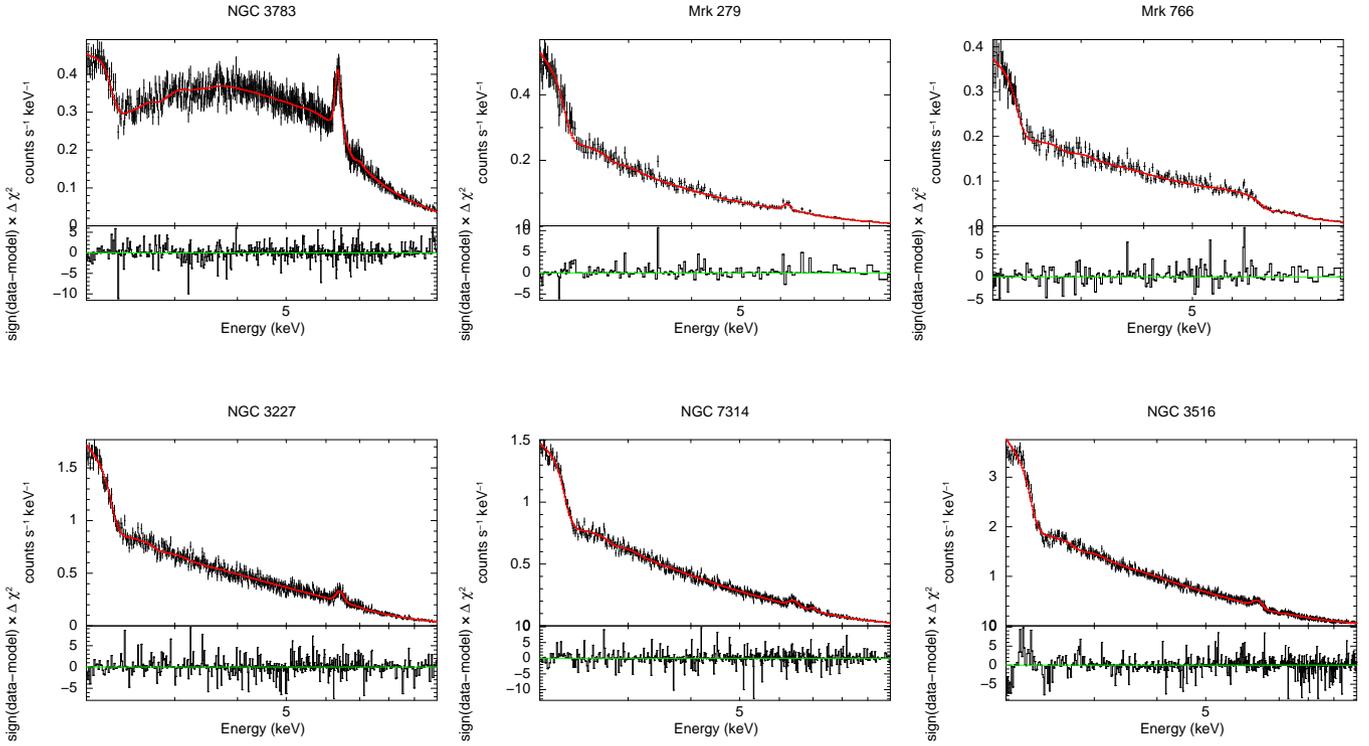

	\begin{subfigure}{0.33\textwidth}
	\includegraphics[trim=0 2.3cm 0 0,width=4.5cm,angle=-90]{NGC3783_inter1.ps}
	\vspace{0.3cm}
	\end{subfigure}
	\begin{subfigure}{0.33\textwidth}
	\includegraphics[trim=0 2.3cm 0 0,width=4.5cm,angle =-90]{Mrk279_inter1.ps}
	\vspace{0.3cm}
	\end{subfigure}
	\begin{subfigure}{0.33\textwidth}
	\includegraphics[trim=0 2.3cm 0 0,width=4.5cm,angle =-90]{Mrk766_inter1.ps}
	\vspace{0.3cm}
	\end{subfigure}
	\par\bigskip
	\begin{subfigure}{0.33\textwidth}
	\includegraphics[trim=0 2.3cm 0 0,width=4.5cm,angle =-90]{NGC3227_inter1.ps}
	\vspace{0.3cm}
	\end{subfigure}
    \begin{subfigure}{0.33\textwidth}
	\includegraphics[trim=0 2.3cm 0 0,width=4.5cm,angle =-90]{NGC7314_inter1.ps}
	\vspace{0.3cm}
	\end{subfigure}
	\begin{subfigure}{0.33\textwidth}
	\centering
	\includegraphics[trim=0 2.3cm 0 0,width=4.5cm,angle=-90]{NGC3516_inter1.ps}
	\vspace{0.3cm}
	\end{subfigure}
	\par\bigskip
    \caption{The top panels in each plot show spectra for our sample of AGNs together with the best-fit models (red line) for the first time-interval indicated in Table~\ref{table1}. The bottom panels in each plot show the contribution to the $\chi^2$ value with sign according to the difference of the data and the model for each data point.}\label{figure1}
\end{figure*}

For this analysis, we have used spectra extracted towards the same central regions used for the temporal analysis. A sample of the spectra corresponding to the first time-interval for each galaxy is shown in \mbox{Fig. \ref{figure1}}.

Using Xspec v12.12.0 \citep{1996ASPC..101...17A}, we modeled the spectra considering a partial absorber with free column density ($N_{\rm H}$) and covering fraction ($C_{\rm F}$), one Gaussian emission line at 6.4 keV and a power-law continuum. The $N_{\rm H}$ and $C_{\rm F}$ components are included in our modeling as we are interested in studying changes in these two parameters of the absorbers that could be eclipsing the central X-ray source.
Our best-fitting models are indicated in Fig. \ref{figure1} and the  estimated parameters are listed in Table \ref{Table2}. The above model allowed us to fit well the spectra of the six galaxies - we obtained a reduced $\chi^2$ within $\sim$0.9-1.2. An additional Gaussian line is needed to fit the data of NGC 3783, which is fixed at 6.15 keV and has an equivalent width (EW) of $\sim$250 eV. This additional line was also used by \cite{Blustin2002} to fit the broad component of the neutral Fe K$\alpha$ line of NGC 3783. We have a better fit of the spectrum from time-interval 1 for NGC 3783, considering an absorption line at $\sim$6.7 keV with a full width half maximum (FWHM) of 33 eV.
Furthermore, we introduced an absorption feature at $\sim$7.15 keV with an FWHM within 260-310 eV to obtain a better fit of the spectra for the three time-intervals in Mrk 766. 
The presence of absorption features only in NGC 3783 and Mrk 766 may indicate that the absorber in these two galaxies has a lower ionization degree than in the other galaxies included in our study, which is suggested by \cite{2012MNRAS.424.2255W} as an important parameter affecting the X-ray absorption.
On the other hand, an additional Gaussian line at $\sim$7 keV is required to fit the hydrogen-like Fe K$\alpha$ line in the spectra of the Seyfert 2 galaxy NGC 7314. In the next section, we will study the spectra showing changes in the HR curves by considering shorter time-scales.\\

We also modeled the spectra of the three time-intervals for Mrk 279 and Mrk 766 considering the above components plus the pexrav model \citep{Magdziarz95} to account for neutral reflection.
We modeled the spectrum of the three time intervals of both galaxies keeping all the parameters free.
The parameters obtained with this analysis are given in Table \ref{Table22}. We notice that there is very little difference between the parameters listed in Table \ref{Table2} and those obtained using the pexrav component. Therefore, in this paper, we will consider models without the pexrav component, for simplicity.

\begin{figure}
    \centering
    \includegraphics[trim={1.4cm 0 1.4cm 0},width=0.43\textwidth]{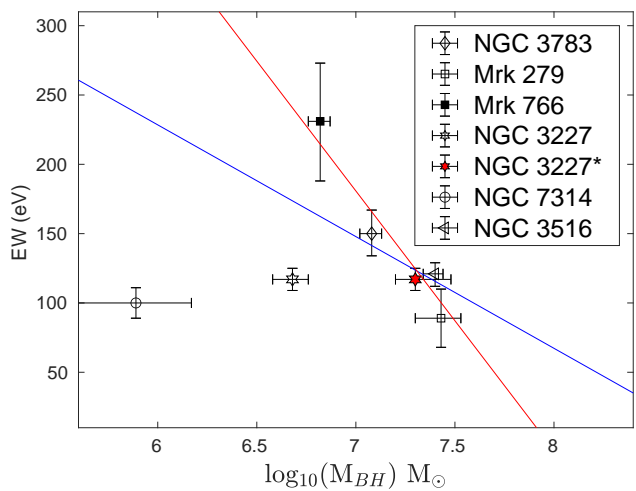}
    \caption{EW of the Fe K$\alpha$ line as function of SMBH mass, for each galaxy. The red line is the fitted line regression obtained considering NGC 3227 with a $\log_{10}(M_{\rm BH})$ of 7.3 $M_{\odot}$ (labeled as NGC 3227* in this figure) and the other four Seyfert 1 galaxies, while the blue line is the fitted line regression obtained considering NGC 3227 with a $\log_{10}(M_{\rm BH})$ of 6.7 $M_{\odot}$ and the other four Seyfert 1 galaxies (see in the text).}\label{Massvsew}
\end{figure}

We have presented the averaged value of the EW of the Fe K$\alpha$ line as a function of the mass of the SMBH in Fig.~\ref{Massvsew}, where the SMBH mass of the Seyfert 1 galaxies and NGC 7314 is taken from \cite{2015PASP..127...67B} and \cite{Emma2016}, respectively. In Fig.~\ref{Massvsew}, we also show a high SMBH mass of 2$\times$10$^7$ $M_{\odot}$ for NGC 3227 (labeled as NGC 3227* in Fig.~\ref{Massvsew}) found by \cite{2008ApJS..174...31H}.
As can be seen in this figure, it appears that the EW is well anti-correlated with the mass of the SMBH when we consider the high SMBH mass for NGC 3227. The Seyfert 1 galaxies in Fig.~\ref{Massvsew} follow a linear decreasing relationship, while the EW value of the Seyfert 2 galaxy (NGC 7314) is an outlier outside this relationship, which will be discussed later. 
To study the dependence between the variables (excluding the Seyfert 2 galaxy), we calculated the Pearson's coefficient and use the least-squares regression taking into account the low (case a) and high (case b) mass estimate of the SMBH for NGC 3227. The Pearson's coefficient was calculated using the PYTHON routine pearsonr of SciPy, while the regression was estimated with the lmfit package\footnote{https://lmfit.github.io/lmfit-py/intro.html}. Thus, we found the following regression for the relation between the EW and the mass of the SMBH in the case a:

\begin{equation}\label{equa1}
EW(eV)=(-80.59\pm33.39)\log(M_{\rm BH} (10^6 M_{\odot}))+(712.00\pm240.95)
\end{equation}

and in the case b:

\begin{equation}\label{equa2}
EW(eV)=(-187.25\pm62.25)\log(M_{\rm BH} (10^6 M_{\odot}))+(1491.54\pm452.98)
\end{equation}

The above relations for the cases a and b are indicated with a blue line and red line, respectively, in Fig. \ref{Massvsew}. We estimated the Pearson's coefficient of -0.48 and -0.87 for the anti-correlation between the logarithm of the SMBH mass and the EW in the case a and b, respectively. The above relations have a p-value of 0.41 in the case a and 0.059 in the case b.
Therefore, there is a better anti-correlation between the two variables in the case b than in the case a.

\subsection{Time-resolved analysis}

\begin{figure*}
    \begin{subfigure}{.3\textwidth}
	\includegraphics[width=1\textwidth,trim=4.2cm 0 4.2cm 0,clip]{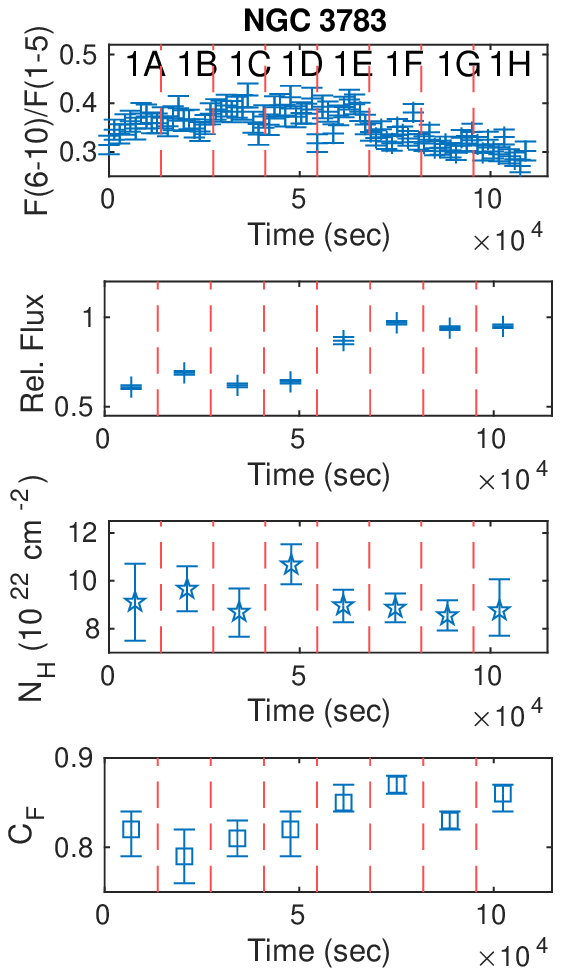}
	\end{subfigure}
	\begin{subfigure}{.3\textwidth}
	\includegraphics[width=1\textwidth,trim=4cm 0 4cm 0,clip]{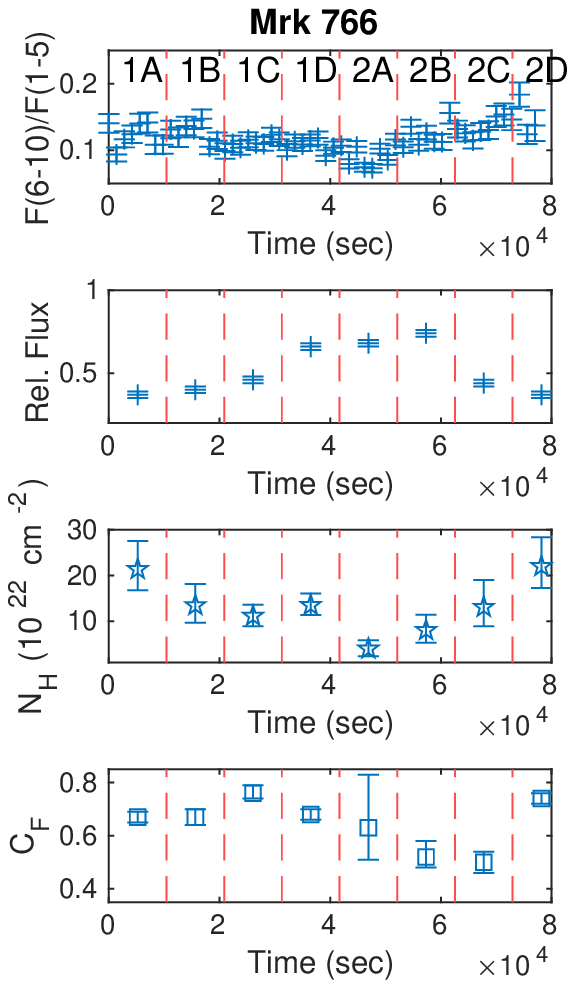}
	\end{subfigure}
	\begin{subfigure}{.3\textwidth}
	\includegraphics[width=1\textwidth,trim=3.6cm 0 3.6cm 0,clip]{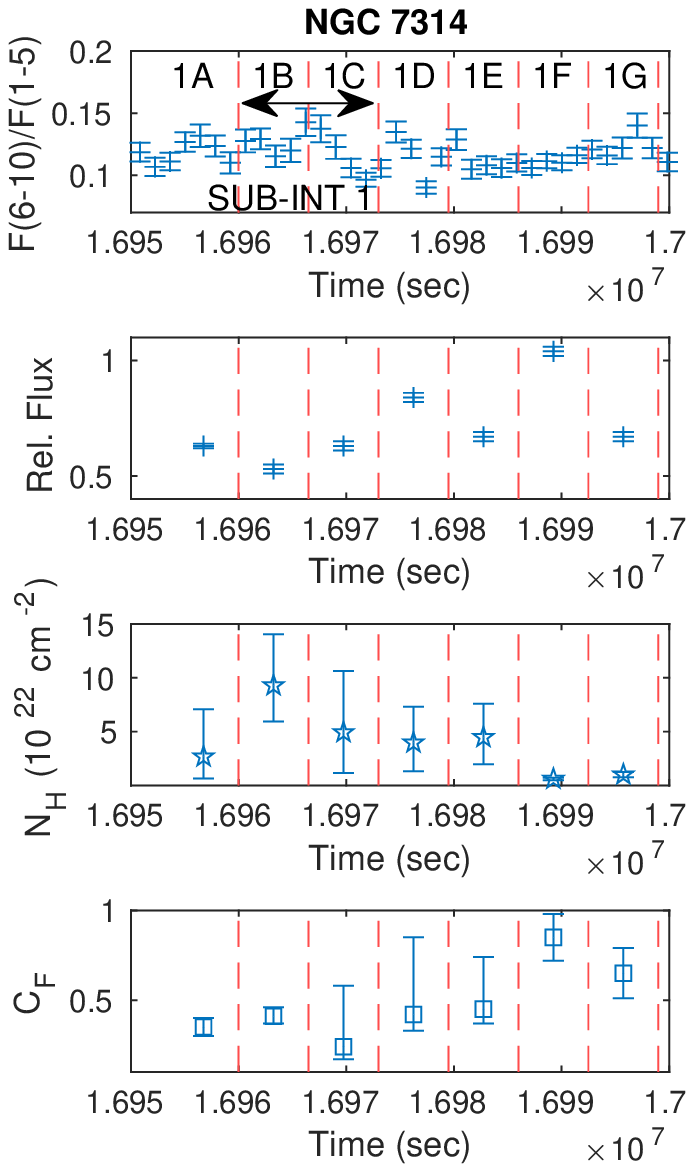}
	\end{subfigure}
	\caption{\textbf{Left panel:} the hardness-ratio, relative flux, column density, and covering factor go from the top to the bottom for NGC 3783 in the first time-interval. The relative flux is the 2-10 keV flux normalized to that estimated in the second interval. \textbf{Middle panel:} as in the left panel but for Mrk 766 in the first time-interval and the relative flux is normalized to that in the third interval. \textbf{Right panel:} as in the left panel but for NGC 7314 in the second time-interval and the relative flux is normalized to that in the first time-interval.}\label{pargalaxy1}
\end{figure*}

\begin{figure*}
    \begin{subfigure}{.3\textwidth}
    \includegraphics[width=1\textwidth,trim=4cm 0 4cm 0,clip]{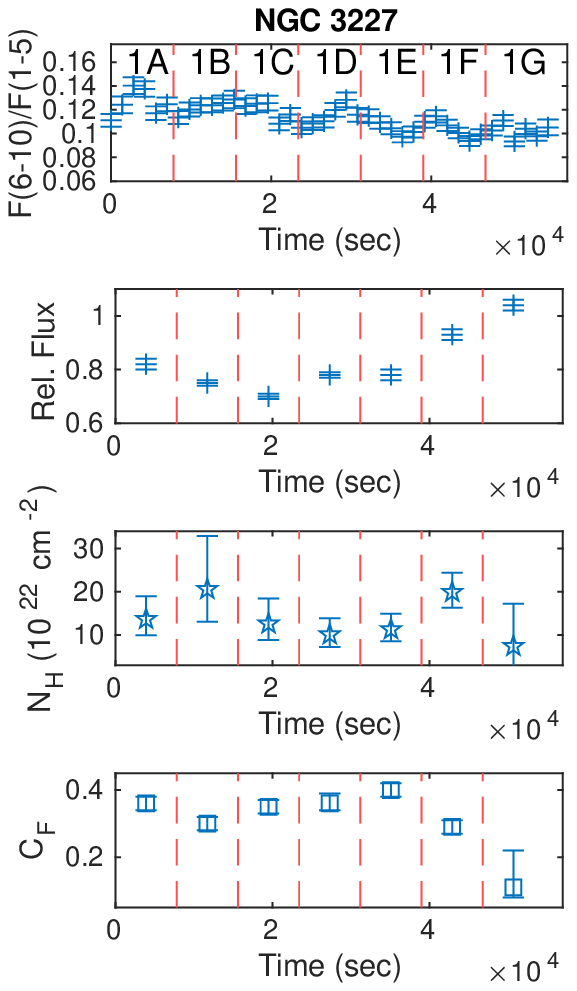}
    \end{subfigure}
	\begin{subfigure}{.3\textwidth}
	\includegraphics[width=1\textwidth,trim=4cm 0 4cm 0,clip]{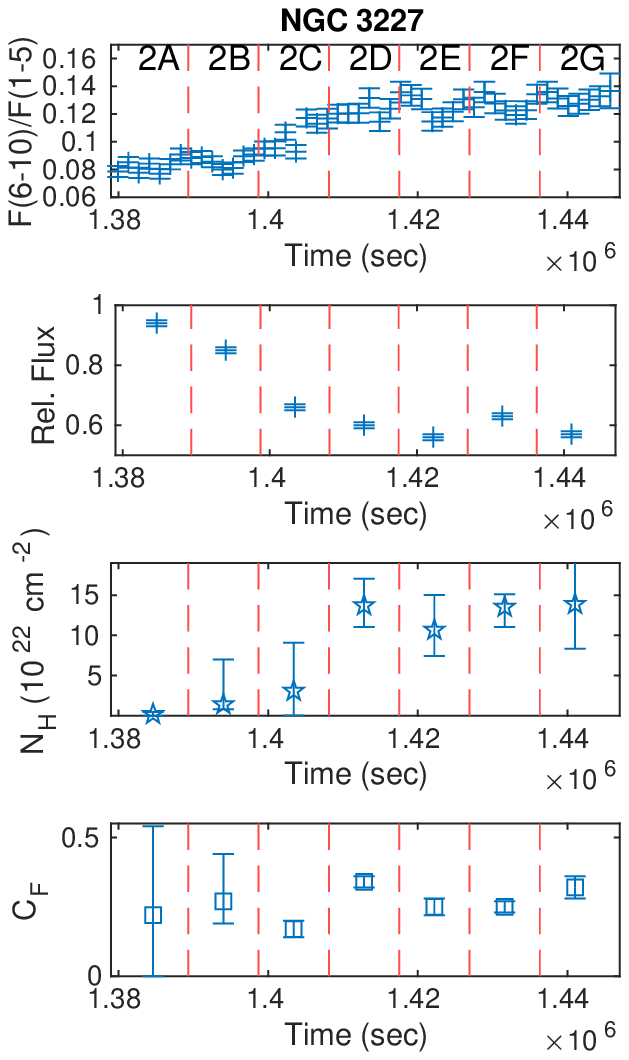}
	\end{subfigure}
	\begin{subfigure}{.3\textwidth}
	\includegraphics[width=1\textwidth,trim=3.8cm 0 3.8cm 0,clip]{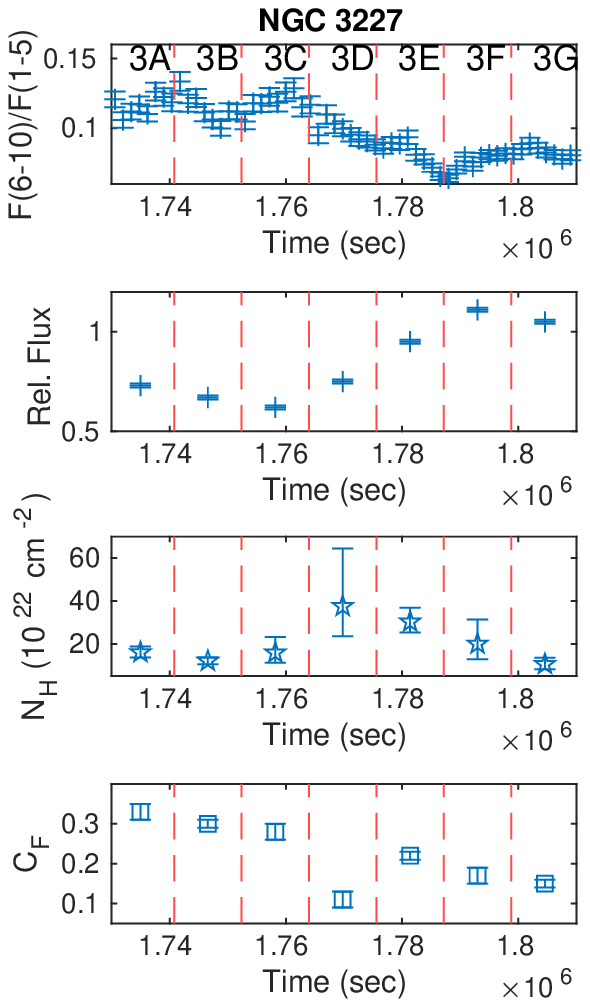}
	\end{subfigure}
	\caption{\textbf{Left panel:} the hardness-ratio, relative flux, column density, and covering factor go from the top to the bottom for NGC 3227 in the first time-interval. The relative flux is the 2-10 keV flux normalized to that estimated in the fourth interval. \textbf{Middle panel:} as in the left panel but for the second time-interval. \textbf{Right panel:} as in the left panel but for the third time-interval.}\label{pargalaxy2}
\end{figure*}

\begin{figure}
    \centering
    \includegraphics[trim=4.3cm 0 4.6cm 0,clip,width=0.3\textwidth]{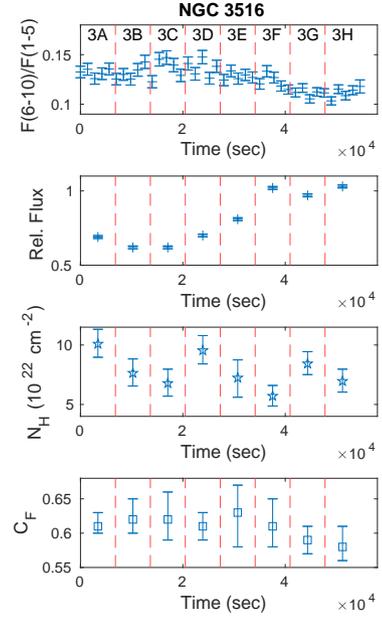}
    \caption{From the top to the bottom: the hardness-ratio, relative flux, column density and covering factor for NGC 3516 in the second time interval. The relative flux is the 2-10 keV flux normalized to that estimated in the fourth time-interval.}\label{parNGC3516}
\end{figure}

In this section, we analyze the spectra of the time-intervals or sub time-intervals showing changes in the HR curves (see Fig. \ref{HRngc3783} and \ref{HRmrk766}-\ref{HRngc3516}) considering short time scales that allow us to identify changes in the relative flux, $N_{\rm H}$ and $C_{\rm F}$. For this, we use uniform time bins and include a number of counts that are large enough for resolving changes in $N_{\rm H}$ and $C_{\rm F}$, using our models. We focus our spectral analysis on the hard X-ray emission (>1 keV) because, as indicated in Section \ref{Analysis1}, NGC 3227 and NGC 3516 do not reveal the occurrence of eclipsing events in the F(0.7-1 keV)/F(0.5-0.7 keV) curves, while the F(6-10 keV)/F(0.7-1 keV) curves do that. Another argument supporting the choice of the hard X-ray range in our analysis is the following: considering energies <2 keV in the spectral analysis leads to values of $N_{\rm H}$<1$\times$10$^{22}$ cm$^{-2}$ and values of $C_{\rm F}$ close to 1 in NGC 3227 and NGC 7314, making it impossible to follow changes in the $C_{\rm F}$ values.
For this analysis, we use the models described in Section \ref{Spectral_analysis} excluding the pexrav component.
The results of our analysis are shown in Table \ref{parameters_sixgalaxies} and Fig. \ref{pargalaxy1}-\ref{parNGC3516}, where we use a number followed by a letter for labeling the time bins. The number of the bin corresponds to that of the time-interval or the sub time-interval. We do not analyze the second time-interval of Mrk 766 because this was already done in \cite{Risaliti2011}.\\

We notice that the first time-interval in Mrk 766 (see Fig.~\ref{pargalaxy1}, middle panel), the first and second time-intervals in NGC 3227 (see Fig.~\ref{pargalaxy2}, left and middle panels), and the third time-interval in NGC 3516  show $N_{\rm H}$ values with a similar trend  to that of the $C_{\rm F}$, but the $N_{\rm H}$ values appear more dispersed (see Fig.~\ref{parNGC3516}). The $N_{\rm H}$ value in the sub time-interval 1F of NGC 3227 is out of the trend (see the left panel in Fig.~\ref{pargalaxy2}), which is not clear.
The $N_{\rm H}$ reveals the highest values in the sub time-intervals 1B and 1C (within SUB-INT 1) of NGC 7314, where we see the change in the HR light curve (see Fig.~\ref{pargalaxy1}, right panel), and the lowest value of $C_{\rm F}$ matches well the lowest value of the HR curve in the SUB-INT 1.
The observed trend between $N_{\rm H}$ and $C_{\rm F}$, in the above time intervals or sub-time intervals of NGC 3227, NGC 7314 and NGC 3516, is consistent with that was found for Mrk 766 by \cite{Risaliti2011}, supporting the idea that during the time intervals that show this trend, a cloud or clouds obscure the center X-ray source.
On the other hand, the values of $N_{\rm H}$ in NGC 3783 show a trend of decreasing with time, while the values of $C_{\rm F}$ reveal a modest increase.
The $N_{\rm H}$ reaches its maximum value around 1.77$\times$10$^6$ seconds and the values of $C_{\rm F}$ show a clear decrease with time in the third time-interval of NGC 3227 (see Fig.~\ref{pargalaxy2}, right panel). It is not clear why the values of $N_{\rm H}$ do not follow the values of the $C_{\rm F}$ in the first time-interval of NGC 3783 and the third time-interval of NGC 3227, which show changes in the HR curves. We also notice in Fig. \ref{pargalaxy1}-\ref{parNGC3516} that there is a clear anti-correlation between the HR curve and the relative flux for the studied galaxies, except for NGC 7314. It is difficult to check changes of the relative flux and of the other two parameters in SUB-INT 1 of NGC 7314, because there are only two possible estimates within this sub time-interval (see right panel of Fig. \ref{pargalaxy1}).

\begin{figure*}
    \begin{subfigure}{.33\textwidth}
    \includegraphics[width=1\textwidth]{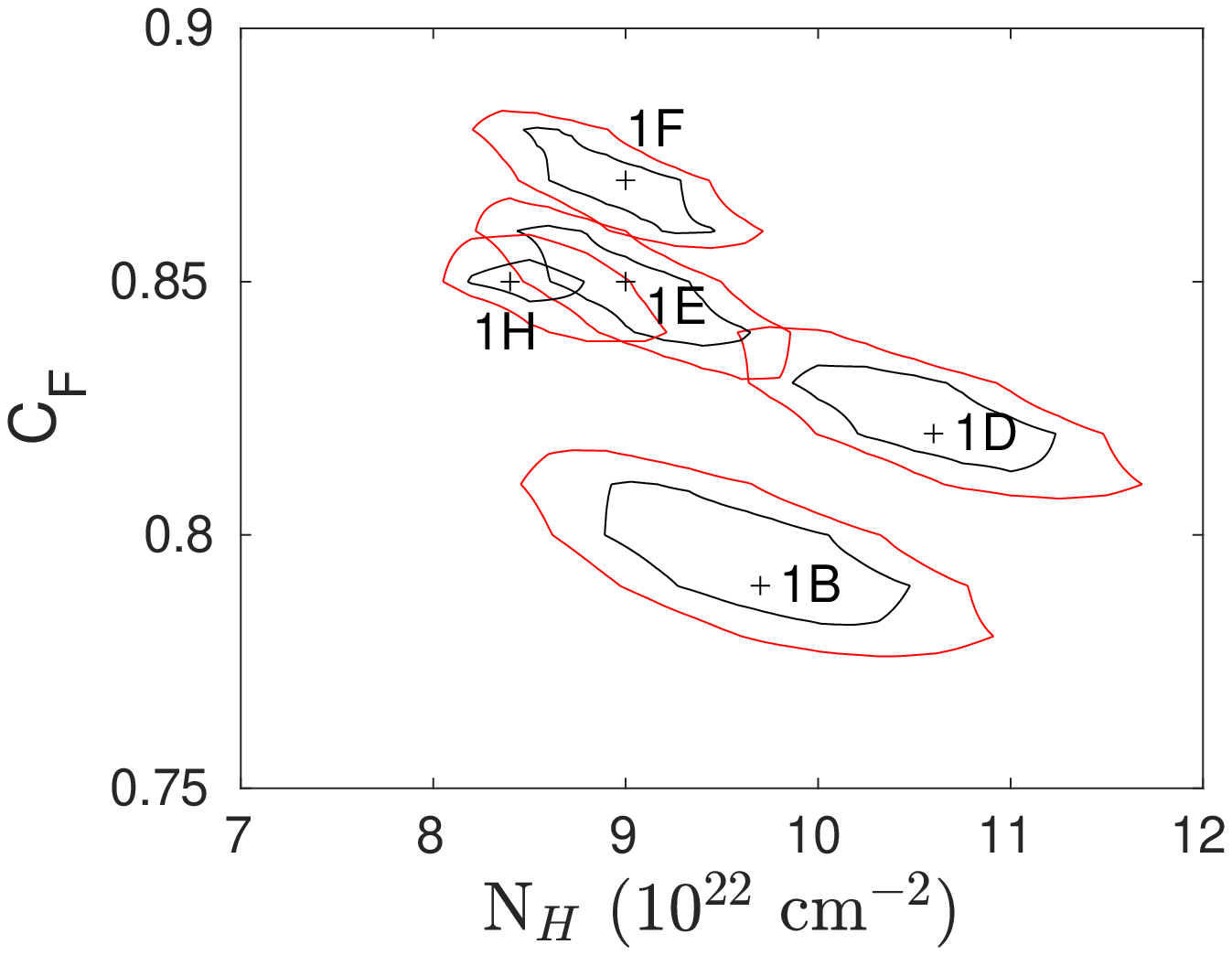}
    \caption{NGC 3783}
    \end{subfigure}
	\begin{subfigure}{.33\textwidth}
	\includegraphics[width=1\textwidth]{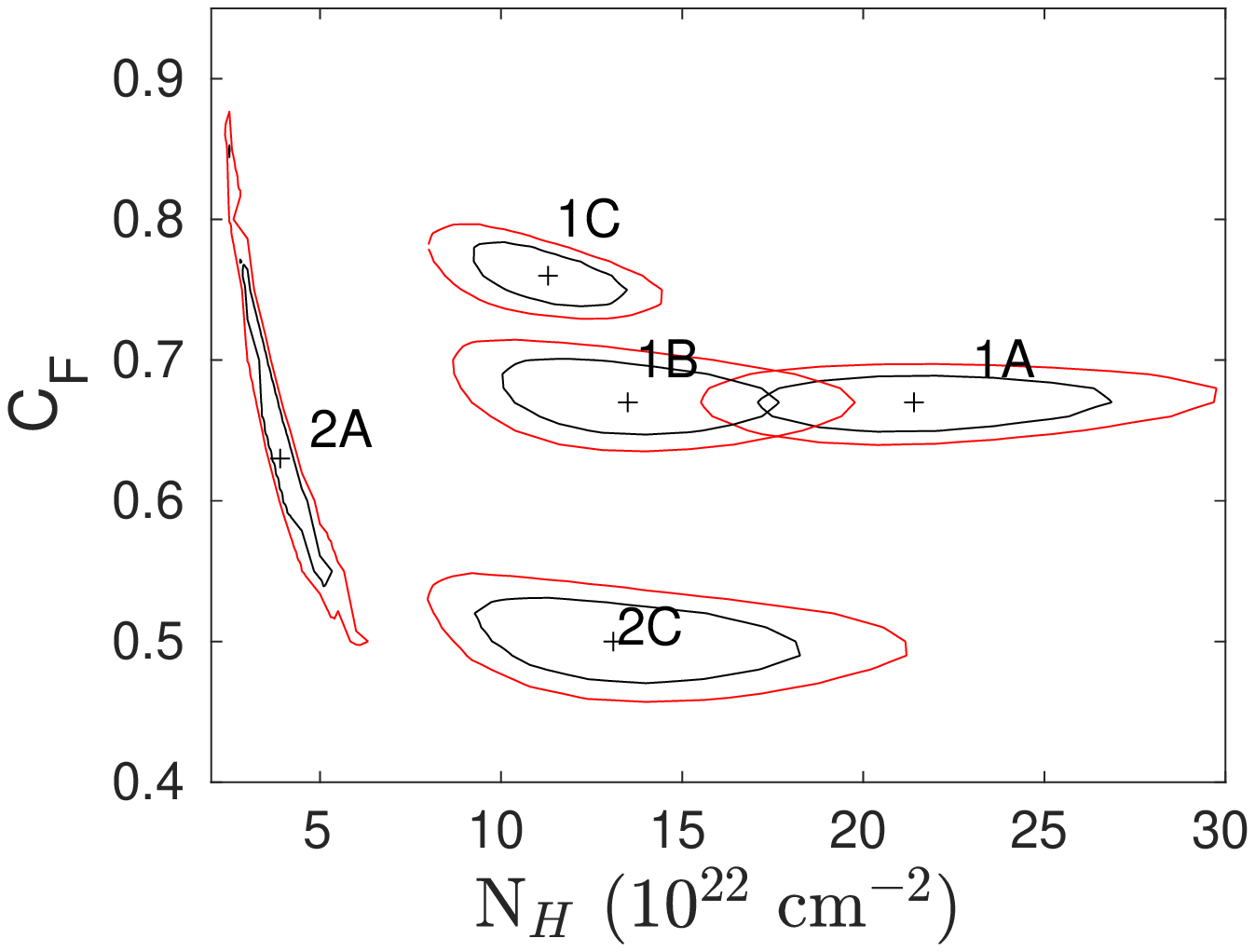}
	\caption{Mrk 766}
	\end{subfigure}
	\begin{subfigure}{.33\textwidth}
	\includegraphics[width=1\textwidth]{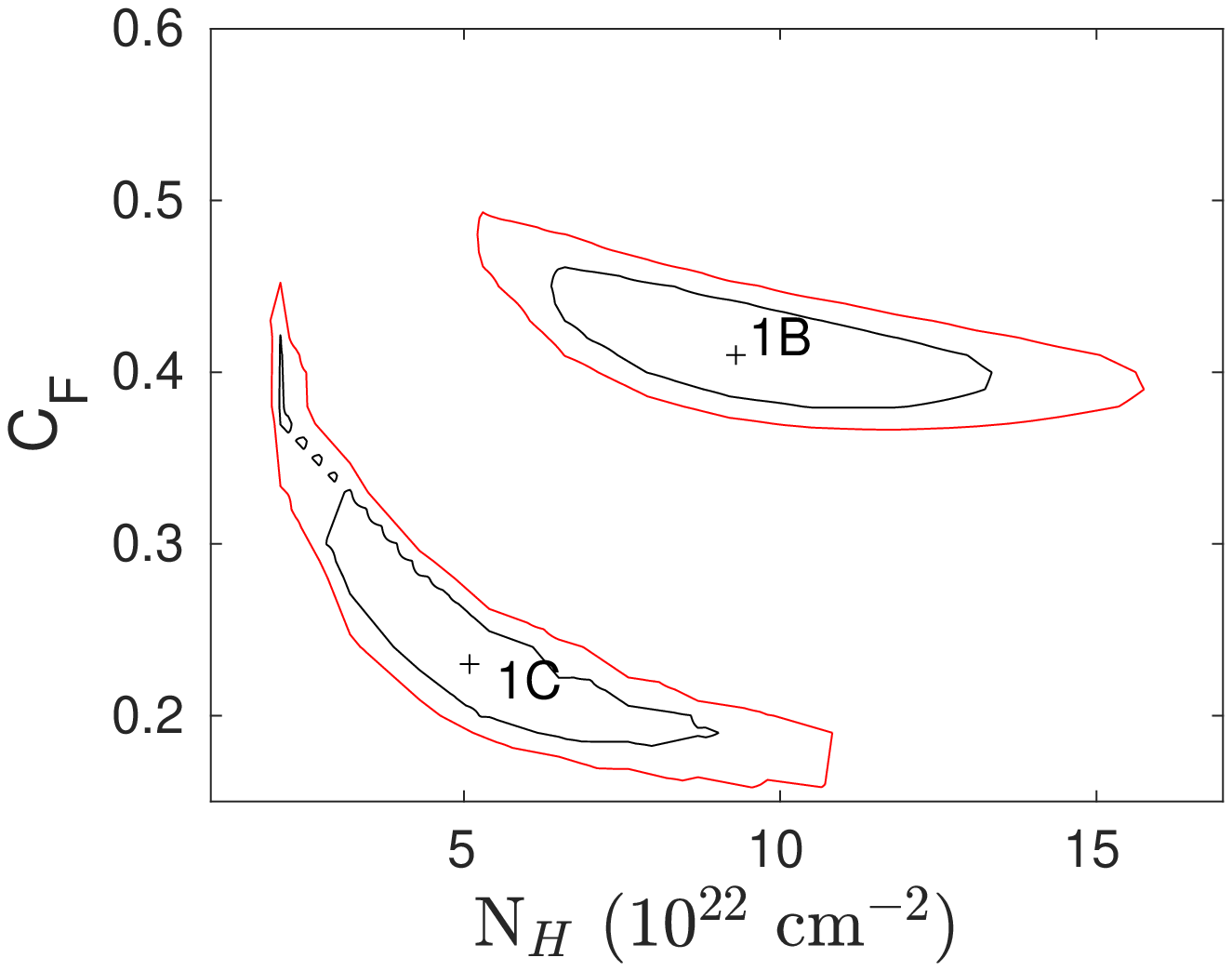}
	\caption{NGC 7314}
	\end{subfigure}
	
	\begin{subfigure}{.33\textwidth}
    \includegraphics[width=1\textwidth]{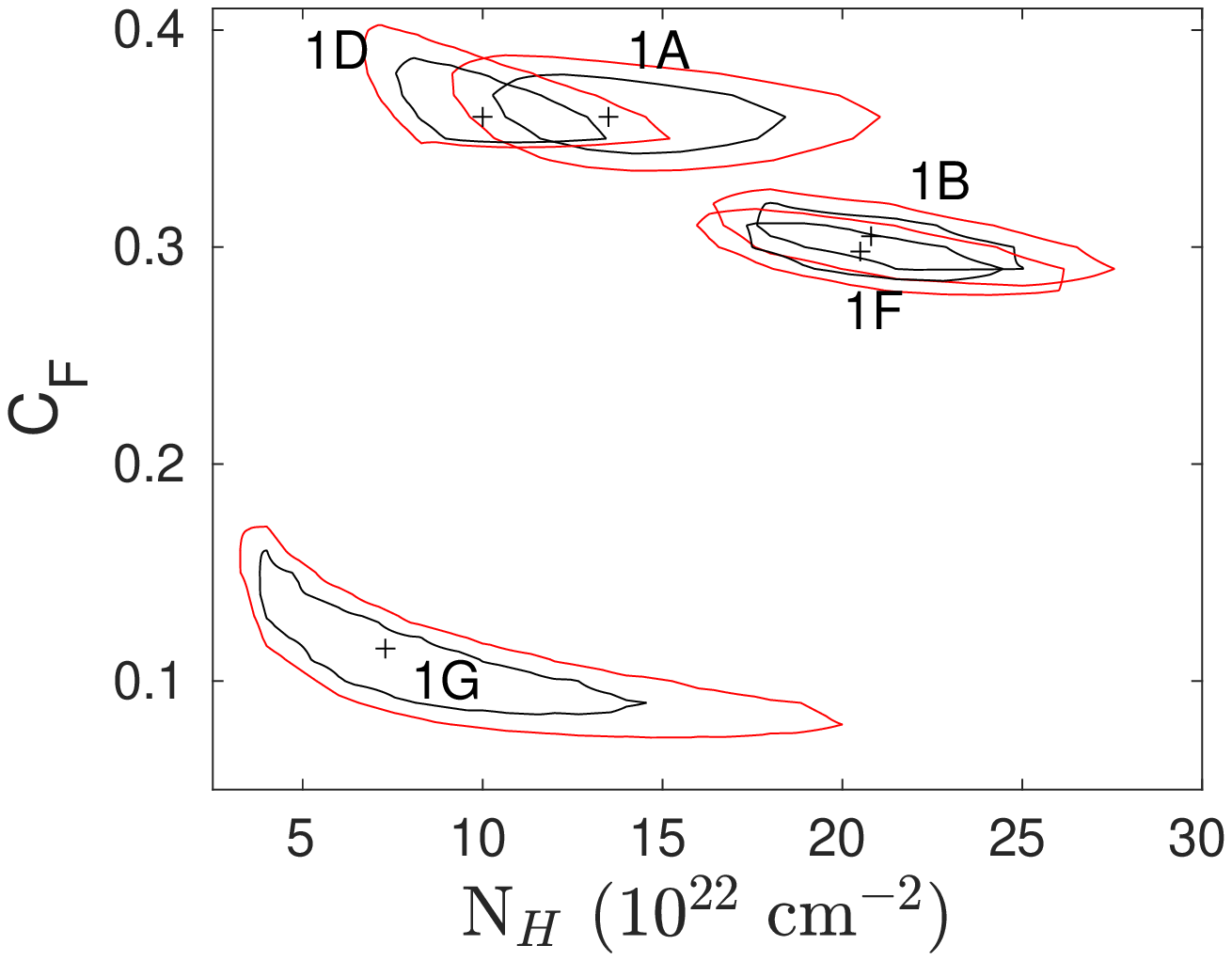}
    \caption{NGC 3227}
    \end{subfigure}
	\begin{subfigure}{.33\textwidth}
	\includegraphics[width=1\textwidth]{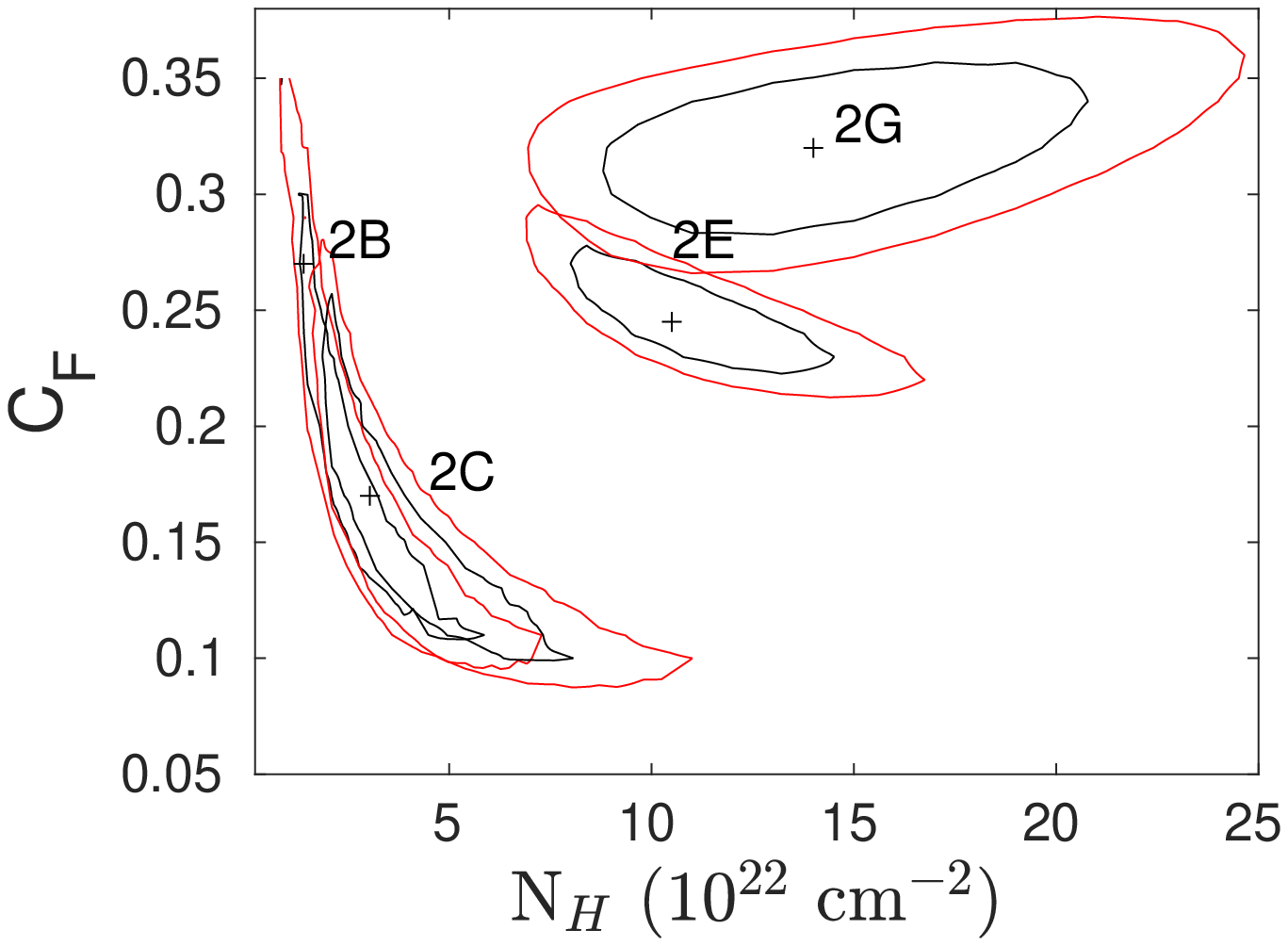}
	\caption{NGC 3227}
	\end{subfigure}
	\begin{subfigure}{.33\textwidth}
	\includegraphics[width=1\textwidth]{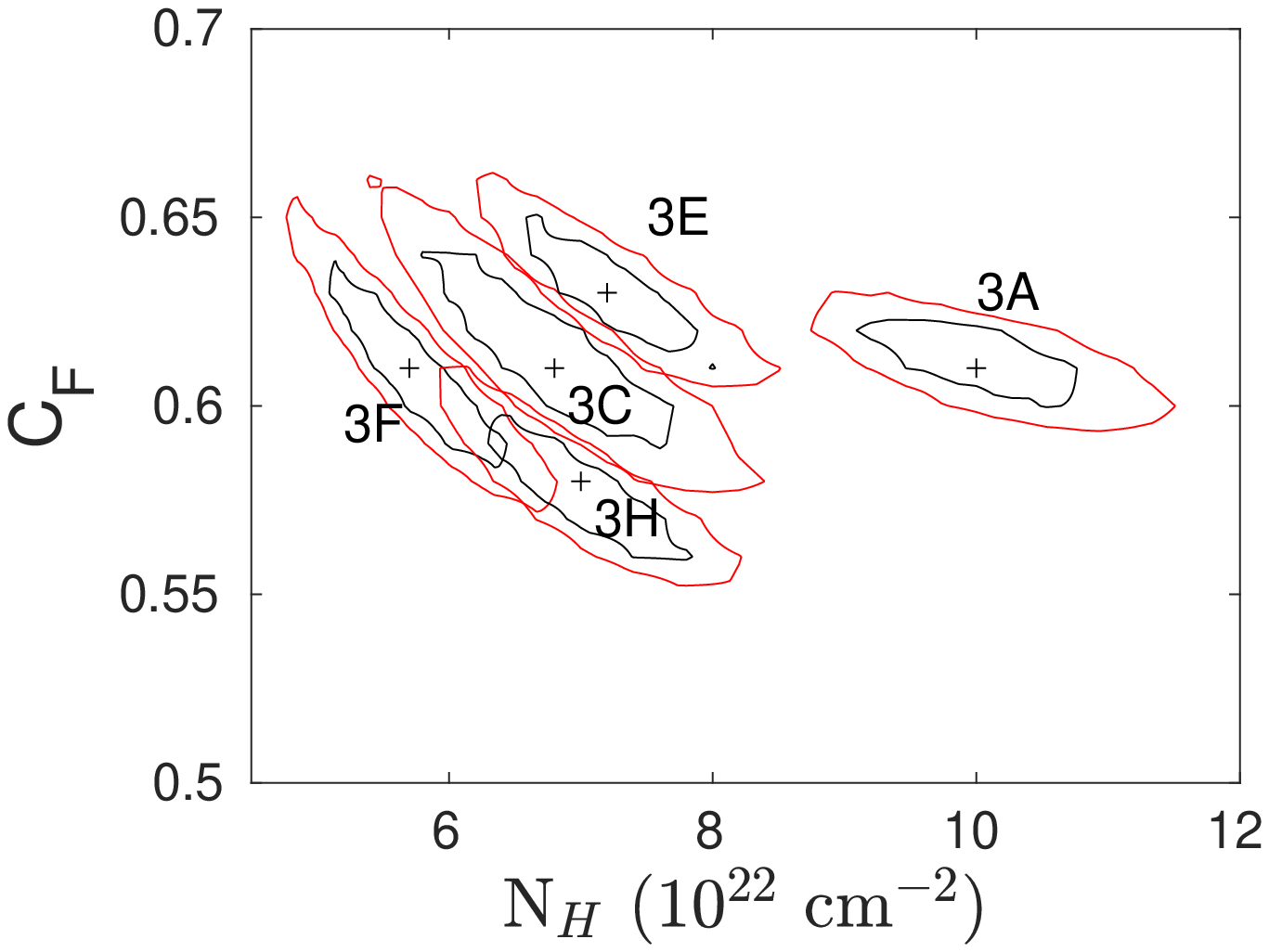}
	\caption{NGC 3516}
	\end{subfigure}
	\caption{Contour plots of $N_{\rm H}$ versus $C_{\rm F}$ for several representative intervals given in Table \ref{parameters_sixgalaxies}. The red contours show the 90 percent confidence level, while the black contours show the 68 percent confidence level. Labels refer to the time sub-intervals as defined in Table \ref{parameters_sixgalaxies}.}\label{NhvsCf_contour}
\end{figure*}

To see the variations in the values of $N_{\rm H}$ and $C_{\rm F}$ with time and their uncertainties, Figure \ref{NhvsCf_contour} shows contour plots of $C_{\rm F}$ versus $N_{\rm H}$ of several representative time intervals given in Table \ref{parameters_sixgalaxies}. We see in this figure that there are uncertainties with overlapping ranges. We estimate these uncertainties with Xspec at 68 and 90 percent confidence levels, which allow us to distinguish the changes in both parameters with time.
For example, the values of $N_{\rm H}$ and $C_{\rm F}$ tend to increase with time in the second time interval of NGC 3227 (panel e in Fig. \ref{NhvsCf_contour}), while the values of $C_{\rm F}$ increase and those of $N_{\rm H}$ decrease with time for the first time interval of NGC 3783 (panel b in Fig. \ref{NhvsCf_contour}).

\section{Discussion}\label{Discussion}

\subsection{Relation between the EW of the 6.4 keV Fe line with the black hole mass}

The best anti-correlation between the values of the EW of the Fe K$\alpha$ line with the SMBH mass is shown in Fig. \ref{Massvsew} for the five Seyfert 1 galaxies studied in this paper, which is fitted using the equation \ref{equa2}. On the other hand, the average EW of the Syfert 2 type NGC 7314 is not consistent with this relation. NGC 7314 with a SMBH mass of 0.9$\times$10$^6$ M$\odot$ reveals an average EW of 100$\pm$11 eV, which is a factor of 2.3 lower than that of Mrk 766 with a SMBH mass of 6.6$\times$10$^6$ M$\odot$. The  average EW for NGC 7314, derived in this contribution, is consistent with the 82$^{+19}_{-21}$ eV found by \cite{2013ApJ...767..121Z} for the same galaxy. 
Considering the linear regression in Fig.~\ref{Massvsew} fitted with equation \ref{equa2}, an EW of $\sim$380 eV is expected for a Seyfert 1 AGN with a SMBH mass of 0.9$\times$10$^6$ M$\odot$ equal to that of NGC 7314.
This contrasts with the findings by \cite{2011A&A...532A..84S}, who found that the EW values of the Fe K$\alpha$ line in Seyfert 2 galaxies are higher than in Seyfert 1 galaxies, which could be a consequence of measuring the EW against the depressed continuum by the torus obscuration in Seyfert 2 galaxies \citep{2011A&A...532A..84S}. On the contrary, our findings are consistent with the attenuation in the luminosity of the Fe K$\alpha$ line in Compton-thick Seyfert 2 galaxies (likely due to absorption of the reflected component) compared to that of Seyfert 1 galaxies with the same mid-IR luminosity \citep{2014MNRAS.441.3622R}.\\

Figure \ref{Massvsew} suggests that the SMBH mass can directly affect the Fe K$\alpha$ line emission region in Seyfert 1 galaxies.
The EW-SMBH mass anti-correlation is expected from the EW-X-ray luminosity anti-correlation known as the X-ray Baldwin effect \citep{1993ApJ...413L..15I,2004MNRAS.347..316P,2010ApJS..187..581S}.
\cite{1997ApJ...477..602N} found that the Fe K$\alpha$ line emission in Seyfert 1 galaxies likely originates in the accretion disk orbiting the SMBH. 
On the other hand, \cite{2011ApJ...727...19F} discovered that the Fe K$\alpha$ line is likely generated in the Compton-thick torus because they found a relation between the EW of the Fe K$\alpha$ line and the absorption column density in Seyfert galaxies.
The accretion rate, proposed as responsible for producing the X-ray Baldwin effect \citep{2009ApJ...690.1322W}, could play an important role in causing the EW-SMBH mass anti-correlation if the Fe K$\alpha$  emission line is generated in the accretion disk.

\subsection{Physical properties of the eclipsing clouds}

We analyzed with the same procedure described by \cite{Risaliti2011} the X-ray variability of six Seyfert galaxies (see Table \ref{table1}), including the galaxy Mrk 766 discussed by \citet{Risaliti2011}. These observations were achieved with the \textit{XMM-Newton} mission using the PN instrument. The AGNs are bright enough [F(2-10) keV] $\gtrsim$ 1$\times$ 10$^{-11}$ erg cm$^{-2}$ s$^{-1}$ with a long observation time of at least $\sim$40 ks. In the same way, we carried out a preliminary analysis of the HR between the high and low-energy light curves [(6-10 keV)$/$(1-5 keV)] to identify intervals where the strongest spectral variations of the X-ray radiation take place due to the occultation by the clouds of the BLR. Then, we identified the occultation events and estimated their durations from the HR light curves.

We can estimate the cloud velocity, having information about the eclipse time and the variation of the covering factor during the occultation, using the expression by \cite{Risaliti2011} derived assuming that the central X-ray source has a size of at least 5$R_g$: $V > 2.5 \times 10^3~ M_6 T^{-1}_4 \sqrt{\Delta C_F} $ [km/s], where $M_6$ is the black hole mass in units of $10^6$ $M_{\odot}$, $T_4$ the occultation time in units of $10^4$ seconds and $\Delta C_F$ is the covering factor variation during the eclipse.
As we will see below, our sources have sizes much larger than five gravitational radii ($R_g = 2\frac{G~ M_{BH}}{c^2}$), so the above expression can be applied undoubtedly.
On the other side, considering that the obscuring clouds are moving with Keplerian velocities, and  combining the transverse velocity of the clouds with their occultation times derived from the variability of the hardness-ratio light curves for each galaxy, a geometrical limit on the X-ray source size can be obtained using the expression given by \cite{2007ApJ...659L.111R}: $D_s = (G M_{BH})^\frac{1}{3} T^\frac{2}{3}$, where $M_{\rm BH}$ is the black hole mass and $T$ is the occultation time. To estimate the size of the eclipsing cloud we consider the distribution of material across the BLR by the covering factor $\Omega\over {4\pi}$, where $\Omega$ is the angle subtended by the clouds, so we use the expression $D_c =4\sqrt{ C_F} D_s$.
 
The physical parameters of the sources and BLR clouds are given in Table \ref{Table5}.
We have not included Mrk 279 in Table \ref{Table5} as this galaxy does not show changes in the HR curves, which is needed for deriving occultation times and variation in the covering factor.
This is the reason why we have not included the parameters of some time-intervals or sub time-intervals of the other AGNs in Table \ref{Table5} either.
The derived values in Table \ref{Table5} for the first and third time-intervals of NGC 3783 and NGC 3227, respectively, may be biased because the $N_{\rm H}$ values do not follow the $C_F$ values, as mentioned above, which is expected when there are gradients in both parameters across the line of sight (see \cite{Risaliti2011}).
It is noted, that at X-ray energies a single cloud can nearly fully block our view of the X-ray source since the cloud is larger than the continuum source (with sizes within $\sim$(30-100) $R_g$). The clouds have linear dimensions of $\sim$10$^{14}$-10$^{15}$ cm, velocities $>$750 km s$^{-1}$, assuming that the clouds orbit the SMBH with Keplerian velocities, and densities $n_c\sim 10^{8}-10^{9}$ cm$^{-3}$. From our analysis, we found that the column density of eclipsing clouds is about $10^{23}$ cm$^{-2}$ (except for NGC 7314 with a column density of 2$\times$10$^{22}$ cm$^{22}$ cm$^{-2}$, see Table \ref{Table2}) and the linear scale of the cloud is of the order of $10^{14}$ cm, then the corresponding density of the cloud is within $10^{8}-10^{9}$ cm$^{-3}$ as was indicated above (see Table \ref{Table5}). This is in agreement with the results by \cite{2012MNRAS.424.2255W}, who found that strong X-ray absorption takes place when eclipsing clouds with column densities of 10$^{22}$-$10^{23}$ cm$^{-2}$ are far from the central black hole, leading to lower ionization degrees and larger opacities of the clouds.
As expected from the expressions used in our estimates (the black hole mass is directly proportional to the velocity and size of the cloud), NGC 3516 with a massive black hole of 25 million solar masses evidences the coupling feedback between the central black hole and the surrounding galaxy, making cloud velocities faster and the size of the clouds and the continuum source larger. Assuming Keplerian velocities, we can estimate distances ($r_{\rm c}$) of the absorbing clouds from the central X-ray source, which are given in the last two columns of Table \ref{Table5}. 

The cloud in NGC 3783 shows a velocity $>$750 km s$^{-1}$ slightly higher than those below 700 km s$^{-1}$ found for the NLR in AGNs \citep{1988ApJS...67..373E,2005MNRAS.357..220R}, while the other AGNs reveal cloud velocities $>$1122 km s$^{-1}$, so we may be observing a NLR cloud instead of a BLR cloud in NGC 3783.
As mentioned above, the cloud velocity derived for NGC 3783 may be biased because the values of $N_{\rm H}$ do not follow those of $C_{\rm F}$ as expected when there are variations in both parameters during a eclipsing event. The high density of 8$\times$10$^7$ cm$^{-3}$ estimated for NGC 3783 supports the idea that the eclipsing cloud in this AGN is likely a BLR cloud. This density is much larger than those of $<10^5$ cm$^{-3}$ found for NLRs \citep{2006A&A...447..863N}.

The changes in HR curves plus the variation trend between the $N_{\rm H}$ values and the $C_{\rm F}$ values in Mrk 766, NGC 3227, NGC 7314, and NGC 3516 evidence that BLR clouds are likely eclipsing the central X-ray sources. The BLR clouds eclipsing the central X-ray source in Mrk 766 were studied in detail in \cite{Risaliti2011}. The occultation events observed in NGC 3516 and NGC 3227 are in agreement with the flux variations that could be due to the passage of clumps across the line of sight in NGC 3516 \citep{2011ApJ...733...48T} and the detection of a transient obscuration event in NGC 3227 \citep{2021A&A...652A.150M}, which would be caused by obscuring winds as opposed to eclipsing clouds. \cite{2011A&A...535A..62E} proposed that neutral gas grazing the clumpy torus in NGC 7314 and crossing our line of sight is responsible for the variations in the absorption properties in NGC 7314, which is consistent with our findings but our study shows that the absorber is located in the BLR. \cite{2020A&A...634A..65D} estimated a density of $>$7$\times$10$^7$ cm$^{-3}$ for the obscuring gas in the BLR of NGC 3783, which is consistent with the density of the eclipsing cloud found for this galaxy. The column density and the distance (from the central X-ray source) of the eclipsing cloud in NGC 3783 agree with those derived by \cite{2004ApJ...602..648R} using the absorption line at 6.67 keV, which was also detected in our study (see Section \ref{Spectral_analysis}).
Moreover, the column density and covering factor estimated for the eclipsing cloud in NGC 3516 are in agreement with those found by \cite{2008A&A...483..161T} for an absorber eclipsing the continuum central source in NGC 3516.

The derived distances $r_{\rm c}$ of the eclipsing clouds from the X-ray source are within (0.3-8)$\times$10$^4$ $R_{\rm g}$ (see Table \ref{Table5}).
The $r_{\rm c}$ value of 8$\times$10$^4$ $R_{\rm g}$ derived for NGC 3783 agrees with those of $(7.4-8.6)\times$10$^4$ $R_{\rm g}$ estimated by \cite{2014MNRAS.439.1403M} for the X-ray absorbing clouds located at the dusty torus of this AGN. 
The $r_{\rm c}$ values within (3-8)$\times$10$^3$ $R_{\rm g}$ derived for Mrk 766 are consistent with those derived by \cite{Risaliti2011} for BLR clouds in this AGN. We find $r_{\rm c}$ values within $\sim$(1-2)$\times$10$^4$ $R_{\rm g}$ for NGC 3227, which are lower than those of (0.7-2)$\times$10$^5$ $R_{\rm g}$ found for X-ray absorbing clouds located in the dusty torus of this galaxy \citep{2014MNRAS.439.1403M}. 
We would obtain lower distances around $10^3$ $R_{\rm g}$ for the eclipsing clouds in NGC 3227 if we consider the high SMBH mass of 2$\times$10$^7$ M$_{\odot}$ used in our analysis in Section \ref{Spectral_analysis}, which are lower than the BLR cloud distances within $\sim$(0.5-7)$\times$10$^4$ $R_{\rm g}$ measured for NGC 3227 \citep{2014MNRAS.439.1403M}.
Furthermore, the $r_{\rm c}$ values of 9.6$\times$10$^{15}$ cm derived for NGC 7314 and of 3.2$\times$10$^{16}$ cm derived for NGC 3516 are consistent with those expected in BLRs of AGNs with a SMBH mass within $\sim$(0.1-2)$\times$10$^7$ M$_{\odot}$  \citep[see][]{2019A&A...628A..26P}.
Thus, the clouds obscuring the central X-ray source in Mrk 766, NGC 3227, NGC 7314, and NGC 3516 show $r_{\rm c}$ values of (0.3-3.6)$\times$10$^4$ $R_{\rm g}$, typical of BLR clouds.

\begin{table*}
	\centering
	\caption{Spectral analysis parameters.}
	\label{Table2}
	\begin{threeparttable}
	\begin{tabular}{llccrcrrr} 
		\hline
		Galaxy & Interval & $\Gamma$\tnote{a} & N$_{\rm H}$\tnote{b} & C$_{\rm F}$\tnote{c} & Energy\tnote{d} & EW\tnote{e} & f(2-10 keV)\tnote{f} & $\chi^2$/d.o.f.\\
		      &  & &  ($\times$10$^{22}$ cm$^{-2}$) &  & (keV) & (eV) & ($\times$10$^{-11}$ erg s$^{-1}$ cm$^{-2}$) &   \\
		\hline
NGC 3783 & 1 & 1.61$\pm$0.02 & 8.61$^{+0.26}_{-0.25}$ & 0.83$\pm$0.01 & 6.42$^{+0.01}_{-0.01}$ & 156$^{+17}_{-20}$ & 2.41$\pm$0.01 & 799/787\\
         & 2 & 1.77$\pm$0.01 & 8.08$^{+0.38}_{-0.37}$ & 0.76$\pm$0.01 & 6.40$^{+0.01}_{-0.01}$ & 144$^{+28}_{-27}$ & 3.00$\pm$0.02 & 760/753\\
         \hline
Mrk 279	 & 1 & 1.97$\pm$0.06 & 33.06$^{+35.33}_{-11.09}$ & 0.20$\pm$0.03 & 6.42$^{+0.02}_{-0.03}$ & 82$^{+20}_{-20}$ & 2.68$\pm$0.02 & 617/657\\
		 & 2 & 1.94$\pm$0.05 & 119.13$^{+59.88}_{-29.51}$ & 0.34$\pm$0.04 & 6.37$^{+0.04}_{-0.04}$ & 102$^{+41}_{-43}$ & 2.35$\pm$0.03 & 416/441\\
		 & 3 & 1.97$\pm$0.06 & 98.60$^{+43.65}_{-22.99}$ & 0.30$\pm$0.02 & 6.40$^{+0.04}_{-0.13}$& 83$^{+43}_{-41}$ & 2.36$\pm$0.04 & 324/344\\
		 \hline
Mrk 766	 & 1 & 1.91$\pm$0.10 &  11.10$^{+1.26}_{-1.13}$ & 0.56$\pm$0.01 & 6.40\tnote{g} & 308$^{+86}_{-89}$ & 0.72$\pm$0.01 & 699/591\\
		 & 2 &  2.18$\pm$0.02 & 10.66$^{+2.98}_{-2.38}$ & 0.30$\pm$0.01 & 6.40\tnote{g} & 180$^{+66}_{-63}$ & 1.11$\pm$0.01 & 555/572\\
         & 3 & 2.15$\pm$0.08 &   6.76$^{+3.42}_{-2.66}$ & 0.19$\pm$0.04 & 6.40\tnote{g} & 206$^{+71}_{-65}$ & 1.40$\pm$0.01 & 602/608\\
         \hline
NGC 3227 & 1 & 1.67$\pm$0.05 & 11.95$^{+1.73}_{-1.51}$ & 0.32$\pm$0.01 & 6.41$^{+0.02}_{-0.02}$ & 117$^{+18}_{-16}$ & 3.13$\pm$0.02 & 795/735 \\
         & 2 & 1.60$\pm$0.06 & 7.70$^{+2.95}_{-2.32}$ & 0.17$\pm$0.02 & 6.41$^{+0.01}_{-0.01}$ & 146$^{+17}_{-17}$ & 2.59$\pm$0.02 & 778/740 \\
         & 3 & 1.70$\pm$0.01 & 16.67$^{+4.05}_{-3.09}$ & 0.19$\pm$0.01 & 6.39$^{+0.02}_{-0.02}$ & 95$^{+15}_{-12}$ & 3.14$\pm$0.01 & 840/769 \\
         & 4 & 1.72$\pm$0.05 & 8.08$^{+2.94}_{-3.54}$ & 0.16$\pm$0.05 & 6.41$^{+0.02}_{-0.01}$ & 111$^{+15}_{-15}$ & 3.69$\pm$0.02& 819/766 \\
         \hline
NGC 7314 & 1 & 1.81$\pm$0.01& 1.35$^{+0.78}_{-0.38}$ & 0.68$\pm$0.02 & 6.44$^{+0.02}_{-0.03}$ & 86$^{+13}_{-14}$ & 2.39$\pm$0.01 & 915/782 \\ 
         & 2 & 1.80$\pm$0.04& 1.98$^{+0.84}_{-0.80}$ & 0.48$\pm$0.22 & 6.42$^{+0.03}_{-0.03}$ & 114$^{+18}_{-17}$ & 1.99$\pm$0.01 & 841/763 \\
		\hline
NGC 3516 & 1 & 2.03$\pm$0.01& 7.37$^{+0.56}_{-0.53}$ & 0.45$\pm$0.01 & 6.39$^{+0.03}_{-0.02}$ & 89$^{+19}_{-20}$ & 5.25$\pm$0.02 & 928/768 \\ 
         & 2 & 2.05$\pm$0.01& 7.90$^{+0.51}_{-0.48}$ & 0.48$\pm$0.01 & 6.37$^{+0.02}_{-0.01}$ & 123$^{+15}_{-13}$ & 4.58$\pm$0.02 & 954/766 \\	
         & 3 & 2.02$\pm$0.01 & 8.07$^{+0.43}_{-0.41}$ & 0.61$\pm$0.01 & 6.38$^{+0.02}_{-0.01}$ & 139$^{+18}_{-16}$ & 3.56$\pm$0.02 & 864/743 \\
         & 4 & 2.05$\pm$0.04 & 7.13$^{+0.73}_{-0.77}$ & 0.48$\pm$0.03 & 6.41$^{+0.17}_{-0.06}$ & 133$^{+17}_{-15}$ & 4.59$\pm$0.02 & 890/769 \\
        \hline
	\end{tabular}
	\begin{tablenotes}
	\item[a] Photon index of the continuum. The uncertainties given in this table are at the 90 per cent confidence level for one parameter of interest.
	\item[b] Column density of hydrogen.
	\item[c] Covering factor.
	\item[d] Peak energy of the Fe K$\alpha$ line.
	\item[e] Equivalent width of the 6.4 Fe K$\alpha$ line.
	\item[f] The 2-10 keV flux.
    \item[g] Fixed parameter.
	\end{tablenotes}
\end{threeparttable}
\end{table*}

\begin{table*}
	\centering
	\caption{Spectral analysis parameters obtained using pexrav.}
	\label{Table22}
	\begin{threeparttable}
	\begin{tabular}{llccrcrrr} 
		\hline
		Galaxy & Interval & $\Gamma$\tnote{a} & N$_{\rm H}$\tnote{b} & C$_{\rm F}$\tnote{c} & Energy\tnote{d} & EW\tnote{e} & f(2-10)\tnote{f} & $\chi^2$/d.o.f.\\
		      &  & &  ($\times$10$^{22}$ cm$^{-2}$) &  & (keV) & (eV) & ($\times$10$^{-11}$ erg s$^{-1}$ cm$^{-2}$) &   \\
		\hline
Mrk 279 & 1 & 1.99$\pm$0.05 & 30.76$^{+22.19}_{-8.55}$ & 0.20$\pm$0.03 & 6.42$^{+0.02}_{-0.03}$ & 80$^{+1}_{-4}$ & 2.68$\pm$0.02 & 616/657\\
        & 2 & 1.89$\pm$0.13 &  111.78$^{+28.26}_{-16.38}$ & 0.40$\pm$0.01 & 6.37$^{+0.04}_{-0.04}$ & 95$^{+23}_{-65}$ & 2.35$\pm$0.03 & 416/436\\
        & 3 & 1.83$\pm$0.50 & 95.02$^{+30.55}_{-18.82}$ & 0.40$\pm$0.01 & 6.38$^{+0.04}_{-0.04}$ & 73$^{+38}_{-36}$ & 2.36$\pm$0.05 & 322/341\\
        \hline
Mrk 766 & 1 & 1.91$\pm$0.06 & 8.03$^{+1.10}_{-0.99}$ & 0.49$\pm$0.02 & 6.40\tnote{g} & 113$^{+91}_{-60}$ & 0.72$\pm$0.01 & 694/592\\
        & 2 & 1.91$\pm$0.02 &  15.48$^{+3.84}_{-2.99}$ & 0.20$\pm$0.01 & 6.40\tnote{g} & 135$^{+45}_{-41}$ & 1.11$\pm$0.01 & 555/570\\
        & 3 & 2.07$\pm$0.08 &  4.07$^{+4.35}_{-3.46}$ & 0.14$\pm$0.01 & 6.40\tnote{g} & 127$^{+100}_{-93}$ & 1.40$\pm$0.02 & 600/607\\
        \hline
	\end{tabular}
	\begin{tablenotes}
	\item[a] Photon index of the continuum. The uncertainties given in this table are at the 90 per cent confidence level for one parameter of interest.
	\item[b] Column density of hydrogen.
	\item[c] Covering factor.
	\item[d] Peak energy of the Fe K$\alpha$ line.
	\item[e] Equivalent width of the 6.4 Fe K$\alpha$ line.
	\item[f] The 2-10 keV flux.
    \item[g] Fixed parameter.
	\end{tablenotes}
	\end{threeparttable}
\end{table*}

\begin{table}
    \scriptsize
	\centering
	\caption{Parameters of the time-resolved spectral analysis.}
	\label{parameters_sixgalaxies}
	\begin{threeparttable}
	\begin{tabular}{llcrcr}
		\hline
		Galaxy & Inter. & N$_{\rm H}$\tnote{a}  & C$_{\rm F}$\tnote{b} & F/F'\tnote{c} & $\chi^2$/d.o.f.\\
		       &        & ($\times$10$^{22}$ cm$^{-2}$) & & & \\
		\hline
		NGC 3783 &1A &  9.12$^{+1.60}_{-1.62}$& 0.82$\pm$0.03 & 0.61$\pm$0.01 & 397/409\\
		&1B &  9.66$^{+0.95}_{-0.93}$& 0.79$\pm$0.04 & 0.69$\pm$0.01 & 400/463\\
		&1C &  8.70$^{+0.98}_{-1.04}$& 0.81$\pm$0.02 & 0.62$\pm$0.01 & 383/389\\
		&1D & 10.67$^{+0.86}_{-0.81}$& 0.82$\pm$0.01 & 0.64$\pm$0.01 & 380/408\\
		&1E &  8.97$^{+0.66}_{-0.70}$& 0.85$\pm$0.02 & 0.87$\pm$0.02 & 529/559\\
		&1F &  8.87$^{+0.60}_{-0.60}$& 0.87$\pm$0.01 & 0.97$\pm$0.01 & 593/565\\
		&1G &  8.55$^{+0.64}_{-0.62}$& 0.83$\pm$0.01 & 0.94$\pm$0.01 & 536/557\\
        &1H &  8.76$^{+1.31}_{-1.05}$& 0.86$\pm$0.02 & 0.95$\pm$0.01 & 507/555\\
		\hline
		
		Mrk 766&1A & 21.37$^{+6.19}_{-4.60}$& 0.67$\pm$0.02 & 0.37$\pm$0.02 & 124/108\\
		&1B & 13.49$^{+4.67}_{-3.80}$& 0.67$\pm$0.03 & 0.40$\pm$0.02 & 109/106\\
		&1C & 11.17$^{+2.45}_{-2.24}$& 0.76$\pm$0.03 & 0.46$\pm$0.02 & 124/121\\
		&1D & 13.52$^{+2.54}_{-2.11}$& 0.68$\pm$0.02 & 0.66$\pm$0.02 & 192/174\\
		&2A &  4.00$^{+1.87}_{-1.63}$& 0.63$\pm$0.26 & 0.68$\pm$0.02 & 230/191\\
		&2B &  8.05$^{+3.40}_{-2.71}$& 0.52$\pm$0.05 & 0.74$\pm$0.02 & 197/194\\
		&2C & 13.11$^{+5.93}_{-4.18}$& 0.50$\pm$0.04 & 0.44$\pm$0.02 & 142/153\\
		&2D & 22.02$^{+6.35}_{-4.73}$& 0.74$\pm$0.02 & 0.37$\pm$0.02 & 118/90\\
		\hline
		
		NGC 3227&1A & 13.68$^{+5.26}_{-3.70}$& 0.36$\pm$0.02 & 0.82$\pm$0.02 & 374/407\\
		&1B & 20.59$^{+12.30}_{-7.51}$& 0.30$\pm$0.02 & 0.75$\pm$0.01 & 304/353\\
		&1C & 12.62$^{+5.80}_{-3.76}$& 0.35$\pm$0.02 & 0.70$\pm$0.01 & 371/351\\
		&1D & 10.11$^{+3.80}_{-2.85}$& 0.36$\pm$0.03 & 0.78$\pm$0.01 & 380/371\\
		&1E & 11.31$^{+3.59}_{-2.76}$& 0.40$\pm$0.02 & 0.78$\pm$0.01 & 359/371\\
		&1F & 19.91$^{+4.51}_{-3.61}$& 0.29$\pm$0.02 & 0.93$\pm$0.02 & 371/407\\
		&1G &  7.43$^{+9.80}_{-5.40}$& 0.11$\pm$0.02 & 1.04$\pm$0.02 & 443/434\\
		&2A &  0.18$^{+0.27}_{-0.18}$& 0.22$\pm$0.28 & 0.94$\pm$0.01 & 479/474\\
		&2B &  1.40$^{+5.60}_{-0.60}$& 0.27$\pm$0.13 & 0.85$\pm$0.01 & 456/446\\
		&2C &  3.07$^{+6.01}_{-3.03}$& 0.17$\pm$0.03 & 0.66$\pm$0.01 & 388/394\\
		&2D & 13.72$^{+3.33}_{-2.69}$& 0.34$\pm$0.02 & 0.60$\pm$0.01 & 378/364\\
		&2E & 10.67$^{+4.35}_{-3.23}$& 0.25$\pm$0.03 & 0.56$\pm$0.01 & 335/346\\
		&2F & 13.53$^{+2.58}_{-2.48}$& 0.25$\pm$0.02 & 0.63$\pm$0.01 & 365/374\\
		&2G & 13.88$^{+7.88}_{-5.56}$& 0.32$\pm$0.04 & 0.57$\pm$0.01 & 294/318\\
        &3A & 16.02$^{+2.79}_{-2.35}$& 0.33$\pm$0.02  & 0.73$\pm$0.01 & 471/469 \\
        &3B & 12.00$^{+1.80}_{-1.52}$& 0.30$\pm$0.01  & 0.67$\pm$0.01 & 446/441 \\
        &3C & 15.88$^{+7.33}_{-4.63}$& 0.28$\pm$0.01  & 0.62$\pm$0.01 & 395/427 \\
        &3D & 37.52$^{+27.00}_{-14.00}$& 0.11$\pm$0.02  & 0.75$\pm$0.01 & 416/436 \\
        &3E &30.33$^{+6.49}_{-4.99}$&0.22$\pm$0.01  & 0.95$\pm$0.01 & 557/515 \\
        &3F & 20.03$^{+11.32}_{-7.19}$& 0.17$\pm$0.02& 1.11$\pm$0.01 & 599/545 \\
        &3G & 10.52$^{+3.04}_{-2.27}$& 0.15$\pm$0.01  & 1.05$\pm$0.01 & 479/503\\
		\hline
		
		NGC 7314&1A &  2.67$^{+4.41}_{-2.02}$& 0.35$\pm$0.05 & 0.63$\pm$0.01 & 182/203\\
		&1B &  9.28$^{+4.76}_{-3.35}$& 0.41$\pm$0.05 & 0.53$\pm$0.02 & 160/173\\
		&1C &  4.93$^{+5.78}_{-3.78}$& 0.24$\pm$0.20 & 0.63$\pm$0.02 & 220/206\\
		&1D &  3.97$^{+3.35}_{-2.64}$& 0.42$\pm$0.25 & 0.84$\pm$0.02 & 244/261\\
		&1E &  4.47$^{+3.11}_{-2.50}$& 0.45$\pm$0.20 & 0.67$\pm$0.02 & 218/220\\
		&1F &  0.60$^{+0.12}_{-0.12}$& 0.85$\pm$0.13 & 1.04$\pm$0.02 & 288/306\\
		&1G &  1.00$^{+0.24}_{-0.23}$& 0.65$\pm$0.14 & 0.67$\pm$0.02 & 254/217\\
		\hline
		
		NGC 3516&3A & 10.09$^{+1.23}_{-1.12}$& 0.61$\pm$0.02 & 0.69$\pm$0.01 & 496/456\\
		&3B &  7.64$^{+1.19}_{-1.09}$& 0.62$\pm$0.03 & 0.62$\pm$0.01 & 429/425\\
		&3C &  6.78$^{+1.19}_{-1.08}$& 0.62$\pm$0.04 & 0.62$\pm$0.01 & 437/421\\
		&3D &  9.54$^{+1.25}_{-1.13}$& 0.61$\pm$0.02 & 0.70$\pm$0.01 & 489/456\\
		&3E &  7.24$^{+1.51}_{-1.63}$& 0.63$\pm$0.05 & 0.81$\pm$0.01 & 441/482\\
		&3F &  5.71$^{+0.89}_{-0.84}$& 0.61$\pm$0.04 & 1.02$\pm$0.01 & 555/532\\
		&3G &  8.43$^{+1.01}_{-0.92}$& 0.59$\pm$0.02 & 0.97$\pm$0.01 & 580/523\\
		&3H &  6.96$^{+1.01}_{-0.92}$& 0.58$\pm$0.03 & 1.03$\pm$0.01 & 523/535\\
		\hline
	\end{tabular}
	\begin{tablenotes}
	\item[a] Column density of hydrogen. The uncertainties given in this table are at the 90 per cent confidence level for one parameter of interest.
	\item[b] Covering factor.
    \item[c] Relative flux, which is the 2-10 keV flux normalized to that estimated in the second, third, first, fourth, and fourth time-intervals for NGC 3783, Mrk 766, NGC 3227, NGC 7314 and NGC 3516, respectively.  
    \item Note: Mrk 279 does not show changes in the HR curves, so it does not appear in this table.
	\end{tablenotes}
	\end{threeparttable}
\end{table}

\begin{table*}
	\centering
	\caption{Derived physical parameters.}
	\label{Table5}
	\begin{threeparttable}
	\begin{tabular}{ccrrrrrrrrrr} 
		\hline
		Galaxy & Interval or & M$_{BH}$\tnote{a} & $T_4$\tnote{b} & $\Delta C_{\rm F}$\tnote{c} & V$_{\rm c}$\tnote{d} & D$_{\rm s}$\tnote{e} & D$_{\rm c}$\tnote{f} & n$_{\rm c}$\tnote{g} & r$_{\rm c}$\tnote{h} & r$_{\rm c}$\tnote{i}\\
		      & sub-interval &($\times$10$^{6}~M_{\odot}$) & ($\times$10$^4 s$) &  & (km s$^{-1}$) & ($\times$10$^{13}$ cm) & ($\times$10$^{13}$ cm) & ($\times 10^{8}$cm$^{-3}$) & ($\times$10$^{15}$ cm) & (10$^4$ R$_{\rm g}$)\\
		\hline
NGC 3783 & 1 & 12.0 & 12.0 & 0.09 &  750  & 28.5 & 104.0 & 0.8 & 286.0 & 8.0\\
		 \hline
Mrk 766	 & 1 &  6.6 & 4.0 & 0.34 &  2420 &  11.2 & 33.7 &  3.3 & 15.2 & 0.8\\
		 & 2 &     & 2.6 & 0.32 &  3612 &  8.4 & 18.5 & 5.8 & 6.8 & 0.3\\
         \hline
NGC 3227 & 1 & 4.8 & 5.4 & 0.72 & 1897 & 12.4 & 28.0 & 4.3
& 18.0 & 1.2\\
         & 2 &     & 6.0 & 0.50 & 1423 & 13.3 & 21.9 & 3.5 & 32.0 & 2.2\\
         & 3 &     & 4.3 & 0.49 & 1966 & 10.6 & 18.5 & 9.0 & 16.8 & 1.2\\
         \hline
NGC 7314 & 1 &  0.9& 1.3 & 0.42 & 1122 &  2.7 &  7.6 & 2.6 & 9.6 & 3.6\\
		 \hline
NGC 3516 & 3 & 25.1 & 5.5 & 0.08 & 3227 & 21.7 & 67.7 & 1.2 & 32.3 & 0.4\\ 
         \hline
	\end{tabular}
		\begin{tablenotes}
    \item[a] The SMBH masses are taken from \cite{2015PASP..127...67B} except for NGC 7314 whose SMBH mass is taken from \cite{Emma2016}.
    \item[b] The occultation time.
    \item[c] The covering factor variation during the eclipse.
    \item[d] Velocity of the eclipsing cloud. This value is a lower limit for the velocity.
    \item[e] Size of the central X-ray source.
    \item[f] Size of the eclipsing cloud.
    \item[g] Particle density of the eclipsing cloud.
    \item[h] Distance between the X-ray source and the obscuring cloud.
    \item[i] Distance between the X-ray source and the obscuring cloud in units of R$_{\rm g}$.
	\end{tablenotes}
	\end{threeparttable}
\end{table*}

\section{Conclusions}\label{Conclusion}

We carried out the spectral and temporal analysis of X-ray data of six galaxies observed with the \textit{XMM-Newton} telescope, which was needed for deriving the physical parameters of clouds eclipsing the central X-ray source in five of the six galaxies. Using the hardness-ratio light curves, we identified occultation events towards the central regions of NGC 3783, NGC 3227, NGC 7314, and NGC 3516, as well as corroborated the occultation events in Mrk 766.
The physical size of the central X-ray sources ($\sim$(3-28)$\times$10$^{13}$cm) are less than the size of the eclipsing clouds, thus a single cloud can block the X-ray source and absorb the X-ray spectrum.
The eclipsing clouds in Mrk 766, NGC 3227, NGC 7314, and NGC 3516 have large values of the column densities ($\sim$10$^{22}$-10$^{23}$ cm$^{-2}$) and are located at distances of $\sim$(0.3-3.6)$\times$10$^4$ $R_{\rm g}$, typical of BLR clouds, leading to the notorious temporal variability of the X-ray flux during the time that the cloud is crossing our line of sight. On the other hand, the cloud obscuring the X-ray source in NGC 3783 is likely located in the dusty torus.

We see that the covering factor changes from object to object and the existence of intervening clouds is a common feature in AGNs. The gas in the BLR, located in the vicinity of the black hole is moving at Keplerian velocities of $>$1122 km s$^{-1}$ (excluding the velocities derived for NGC 3783 because toward this source the estimate of the cloud velocities may be biased).

We found a good anti-correlation with a slope of -187$\pm$62 between the known mass of the SMBHs with the EW of the 6.4 keV Fe line for the five Seyfert 1 galaxies considering a $\log_{\rm 10}(M_{\rm BH})$=7.3 M$_{\odot}$ for NGC 3227 in our statistical analysis.
The average value of the EW of NGC 7314, a Seyfert 2 type, does not agree with the SMBH mass-EW relation found for the five Seyfert 1 galaxies, supporting previous results \citep{2014MNRAS.441.3622R}.


\section*{Acknowledgements}
This research was based on observations obtained with \textit{XMM-Newton}, an ESA science mission with instruments and contributions directly funded by ESA Member States and NASA.
We acknowledge the use of the SAS software, developed by ESA's Science Operations Centre staff. This research has used a non-linear least-squares fitting (ttps://lmfit.github.io/lmfit-py/intro.htm), as well as the routine pearsonr of SciPY (https://docs.scipy.org/doc/ scipy/reference/generated/scipy.stats.pearsonr.html).
We thank the anonymous referees for their helpful comments and suggestions that really helped improve the manuscript considerably.

\section*{Data Availability}
The X-ray data used in this study are accessible from the \textit{XMM-Newton} online archive\footnote{http://nxsa.esac.esa.int/nxsa-web/\#search}.



\bibliographystyle{mnras}
\bibliography{mnras_template} 








\bsp	
\label{lastpage}
\end{document}